\documentclass[12pt]{iopart}

\usepackage[utf8]{inputenc}
\usepackage[T1]{fontenc}
\usepackage{iopams} 
\usepackage{graphicx}
\usepackage{tabularx}
\usepackage{booktabs}
\usepackage{pifont}

\usepackage[style=numeric-comp, url=false,isbn=false,doi=true, eprint=false,giveninits=true,maxbibnames=99,maxcitenames=2,date=year,sorting=none,sortcites=true, natbib=true]{biblatex}
\AtEveryBibitem{\clearfield{number}}
\AtEveryBibitem{\clearfield{issue}}
\AtEveryBibitem{\clearfield{pages}}
\AtEveryBibitem{\clearfield{pagetotal}}
\DeclareNameAlias{author}{last-first}
\DeclareFieldFormat{titlecase}{\MakeTitleCase{#1}}
\newrobustcmd{\MakeTitleCase}[1]{%
  \ifthenelse{\ifcurrentfield{booktitle}\OR\ifcurrentfield{booksubtitle}%
    \OR\ifcurrentfield{maintitle}\OR\ifcurrentfield{mainsubtitle}%
    \OR\ifcurrentfield{journaltitle}\OR\ifcurrentfield{journalsubtitle}%
    \OR\ifcurrentfield{issuetitle}\OR\ifcurrentfield{issuesubtitle}%
    \OR\ifentrytype{book}\OR\ifentrytype{mvbook}\OR\ifentrytype{bookinbook}%
    \OR\ifentrytype{booklet}\OR\ifentrytype{suppbook}%
    \OR\ifentrytype{collection}\OR\ifentrytype{mvcollection}%
    \OR\ifentrytype{suppcollection}\OR\ifentrytype{manual}%
    \OR\ifentrytype{periodical}\OR\ifentrytype{suppperiodical}%
    \OR\ifentrytype{proceedings}\OR\ifentrytype{mvproceedings}%
    \OR\ifentrytype{reference}\OR\ifentrytype{mvreference}%
    \OR\ifentrytype{report}\OR\ifentrytype{thesis}}
    {#1}
    {\MakeSentenceCase{#1}}}
\begin{document}

\title[Measuring tropical rainforest resilience under non-Gaussian disturbances]{Measuring tropical rainforest resilience under non-Gaussian disturbances}

\author{Vitus Benson$^{1,2,3}$, Jonathan F. Donges$^{2,4,5}$, Niklas Boers$^{6,7,8}$, Marina Hirota$^{9,10}$, Andreas Morr$^{6,7}$, Arie Staal$^{4,11}$, Jürgen Vollmer$^3$ \& Nico Wunderling$^{2,4,5}$}

\address{
$^1$Biogeochemical Integration, Max Planck Institute for Biogeochemistry, 07745 Jena, Germany
\\$^2$Earth System Analysis, Potsdam Institute for Climate Impact Research (PIK), Member of the Leibniz Association, 14473 Potsdam, Germany
\\
$^3$Institute for Theoretical Physics, University of Leipzig, 04103 Leipzig, Germany
\\
$^4$Stockholm Resilience Centre, Stockholm University, Stockholm, SE-10691, Sweden
\\
$^5$High Meadows Environmental Institute, Princeton University, Princeton, 08544 New Jersey, USA
\\
$^6$Earth System Modelling, School of Engineering and Design, Technical University of Munich, Munich, Germany
\\
$^7$Complexity Science, Potsdam Institute for Climate Impact Research (PIK), Member of the Leibniz Association, 14473 Potsdam, Germany
\\
$^8$Department of Mathematics and Global Systems Institute, University of Exeter, Exeter, UK
\\
$^9$Department of Physics, Federal University of Santa Catarina, Florianopolis 88040-900-SC, Brasil
\\
$^{10}$Department of Plant Biology, University of Campinas, Campinas 13083-970-SP, Brasil
\\
$^{11}$Copernicus Institute of Sustainable Development, Utrecht University, Utrecht, 3584 CB, The Netherlands
\\
}
\ead{vbenson@bgc-jena.mpg.de, nico.wunderling@pik-potsdam.de}
\vspace{10pt}
\begin{indented}
\item[]October 2023
\end{indented}

\begin{abstract}
    The Amazon rainforest is considered one of the Earth's tipping elements and may lose stability under ongoing climate change. Recently a decrease in tropical rainforest resilience has been identified globally from remotely sensed vegetation data. However, the underlying theory assumes a Gaussian distribution of forest disturbances, which is different from most observed forest stressors such as fires, deforestation, or windthrow. Those stressors often occur in power-law-like distributions and can be approximated by $\alpha$-stable Lévy noise. Here, we show that classical critical slowing down indicators to measure changes in forest resilience are robust under such power-law disturbances. To assess the robustness of critical slowing down indicators, we simulate pulse-like perturbations in an adapted and conceptual model of a tropical rainforest. We find few missed early warnings and few false alarms are achievable simultaneously if the following steps are carried out carefully: First, the model must be known to resolve the timescales of the perturbation. Second, perturbations need to be filtered according to their absolute temporal autocorrelation. Third, critical slowing down has to be assessed using the non-parametric Kendall-$\tau$ slope. These prerequisites allow for an increase in the sensitivity of early warning signals. Hence, our findings imply improved reliability of the interpretation of empirically estimated rainforest resilience through critical slowing down indicators.
    
\end{abstract}

%
%
%
%
%

\section{Introduction}
The Amazon rainforest is considered a crucial component of the Earth's climate system \cite{mitchard_2018} and has been suggested as an Earth system tipping element \cite{armstrongmckay.etal_2022,boers.etal_2022,lenton.etal_2008}. 
There is growing concern that various anthropogenic stressors, such as climate change and associated changes in rainfall patterns, fires, land-use change and deforestation, cause a decrease in resilience and could ultimately lead to large-scale shifts in the Amazons' ecosystem, with severe consequences for the biosphere and human societies \cite{nobre.etal_2016,davidson.etal_2012,boulton.etal_2022,lovejoy.nobre_2018}. Based on conceptual models and observational data, it is believed that the rainforest exhibits the potential for multi-stability at specific levels of moisture supply \cite{hirota.etal_2011,staver.etal_2011,ciemer.etal_2019,staal.etal_2020,wunderling.etal_2022c}. 
This means that certain regions of the rainforest may transition from a rainforest to a savanna-like vegetation state if local precipitation rates are decreased below critical thresholds.

Tropical rainforests, such as the Amazon basin, are subject to multiple stressors. These can originate due to climate (e.g. droughts, heat waves, windthrows), hydrology (e.g. landslides), biotic factors (e.g. insect outbreaks) or anthropogenic activity (e.g. deforestation, wildfires). Many such stressors cause disturbance events, which in turn change the forests in a pulse-like manner. The consequences of disturbances are often visible in the canopy gap structure, i.e. single disturbance events destroy destroy entire parts of the forest, while neighboring parts are almost completely undisturbed. This gap structure has been observed to follow power-law-like distributions \cite{fisher.etal_2008,taubert.etal_2018,gloor.etal_2009,espirito-santo.etal_2010, negron-juarez.etal_2010, asner.etal_2013, chambers.etal_2013, espirito-santo.etal_2014, reis.etal_2022}. In other words, there is a scale-free nature to the gaps, and very large gaps are likely. These power-law-like distributions have also been observed directly for both droughts \cite{stocker.etal_2019} and wildfires \cite{nicoletti.etal_2023} in different ecoregions worldwide. Hence, if we understand disturbances as random perturbations to the forest, i.e. as noise, this noise might have non-Gaussian characteristics. For instance, the noise could be heavy-tailed (e.g. power-law tails). Thus, extreme events become more likely than under Gaussian white noise. Furthermore, Gaussian noise would lead to continuous forest state evolutions, whereas noise due to disturbance events would result in discrete jumps in the time series. Of course, describing all disturbances by one abstract probability distribution is a highly reductive approach given the complexity of the system, yet if it were to be done, a non-Gaussian distribution should not be ruled out.

The resilience of tropical rainforests is usually measured as the response to disturbances directly from observed time series data. The recovery rate to disturbances is related to the temporal autocorrelation of a time series. If a forest is resilient and quickly recovers, the autocorrelation is lower compared to a higher autocorrelation of a vulnerable forest that slowly recovers. Increased autocorrelation in tropical rainforests worldwide has been found using different remotely-sensed vegetation proxies such as above-ground biomass (accounting for water stress, deforestation and vapor-pressure deficit, \cite{saatchi.etal_2021}), vegetation optical-depth (\cite{boulton.etal_2022} only Amazon basin, and \cite{smith.etal_2022} globally), normalized difference vegetation index (NDVI) \cite{lenton.etal_2022} and kernel NDVI \cite{forzieri.etal_2022}. 

Interpreting the autocorrelation as an indicator of resilience requires certain mathematical assumptions \cite{liu.etal_2019a, krakovska.etal_2023}, under which the autocorrelation can also serve as an early warning signal in the approach to a critical transition \cite{vannes.scheffer_2007}. The underlying phenomenon, critical slowing down (CSD), traces a decreasing ability of a system to recover from perturbations when it loses stability. Typically, Gaussian noise is assumed. This is because an analytical relationship between the classical CSD indicators variance and temporal autocorrelation of lag one and the recovery rate $\lambda$ of the dynamics linearized around a given equilibrium can be established \cite{boettner.boers_2022, morr.boers_2023a} (see below). However, as it is not clear that the disturbances occurring in tropical rainforests can be described with Gaussian noise, it is also not clear if the forest resilience can be measured via the autocorrelation.

In this paper, we study forest resilience indicators in the non-Gaussian noise case. For this, we consider $\alpha$-stable Lévy noise, a heavy-tailed generalization of the Gaussian distribution. Depending on the parameters chosen, the noise time series contains jumps (i.e. discrete disturbances events) and the tails follow a power-law (i.e. extreme events become more likely). These characteristics are similar to the ones we postulate for disturbances in tropical rainforests. Hence, it is necessary to understand whether the CSD-based resilience indicators still perform well for Lévy noise. Additionally, we look at pink noise, which contains power-law tails, but no jumps. Previous work has already looked at red noise and time-correlated noise \cite{dutta.etal_2018, boettner.boers_2022, morr.boers_2023a}. Of course, our approach of employing 1D non-Gaussian noise distributions is reductive, as it can not represent any temporal and spatial patterns \cite{bastiaansen.etal_2020, rietkerk.etal_2021}. Yet, since the presence of non-Gaussian noise is plausible for tropical rainforests, the correct functioning of resilience indicators and early warning signals before critical transitions should be assessed thoroughly.

For this assessment, we use a simple conceptual tipping element model of a tropical rainforest \cite{vannes.etal_2014}. This model can represent a real tropical rainforest at the highest level of abstraction, capturing only the stabilising and destabilising feedbacks of the spatially extended system. It gives a good first-order approximation to the tipping structure \cite{vannes.etal_2014, staal.etal_2015}, i.e. the bi-stability of tropical rainforests between rainforest and savannah states \cite{hirota.etal_2011}. For our analysis of forest resilience, it therefore constitutes a good starting point. The argument is a similar one as before: if the indicators do not work in such a simple setting, it is unlikely they would work if real rainforests were to be observed or modelled in more detail.

\begin{figure}
    \centering
    \includegraphics[width=\linewidth]{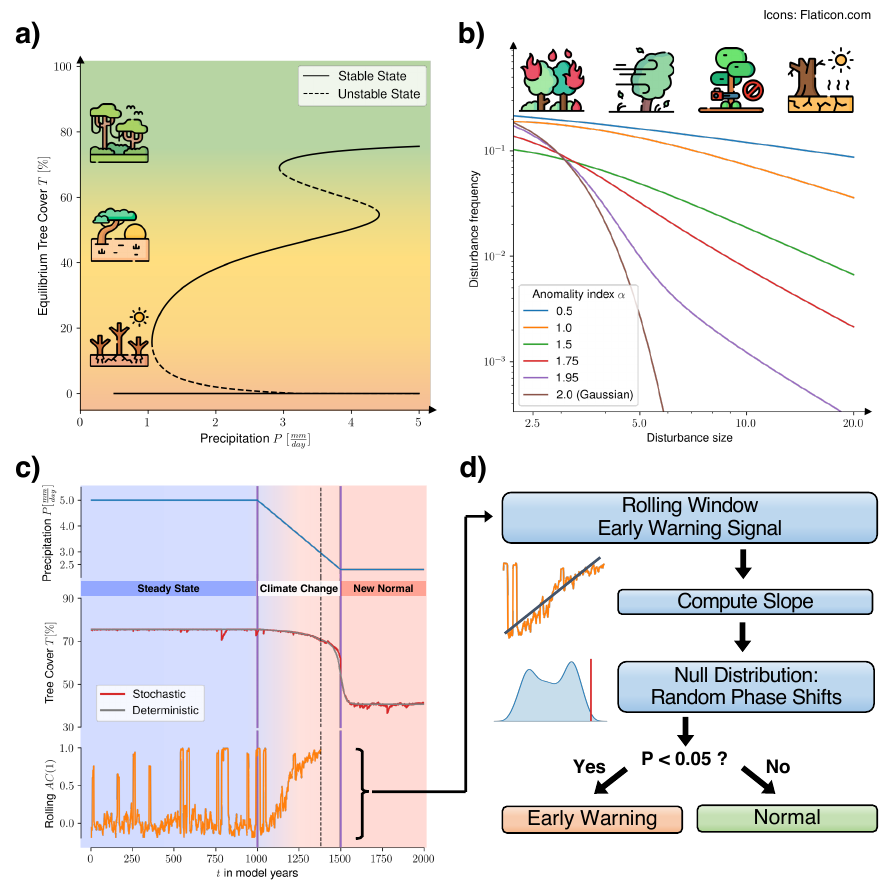} 
   \caption{Overview of the study design. \textbf{a)} Tropical rainforest model bifurcation diagram: Depending on the precipitation level, three different equilibrium states may exist, namely rainforest, savannah and desert. \textbf{b)} Many natural stressors can be modeled through $\alpha$-stable Lévy noise, i.e. in contrast to Gaussian noise ($\alpha = 2$), the tails have power-law decay and hence large events can happen. \textbf{c)} We simulate a climate change scenario, during which we force the precipitation level to decrease below the critical threshold of the rainforest-savannah transition. We compute the rolling autocorrelation $AC(1)$ before the critical transition and observe increases near single large disturbance events and in approach to the critical transition. \textbf{d)} The critical slowing down indicator time series (in this case: $AC(1)$) gets converted into a binary early warning indicator. For this, first a slope is calculated in a window approaching the critical transition. Then, the statistical significance of this slope is assessed by comparing to a null distribution of random phase shift surrogates with a one-sided test. If the slope is significant, there is early warning (i.e. decreasing rainforest resilience).}
    \label{fig:fig1}
\end{figure}

\begin{figure}[t]
    \centering
    \includegraphics[width=\linewidth]{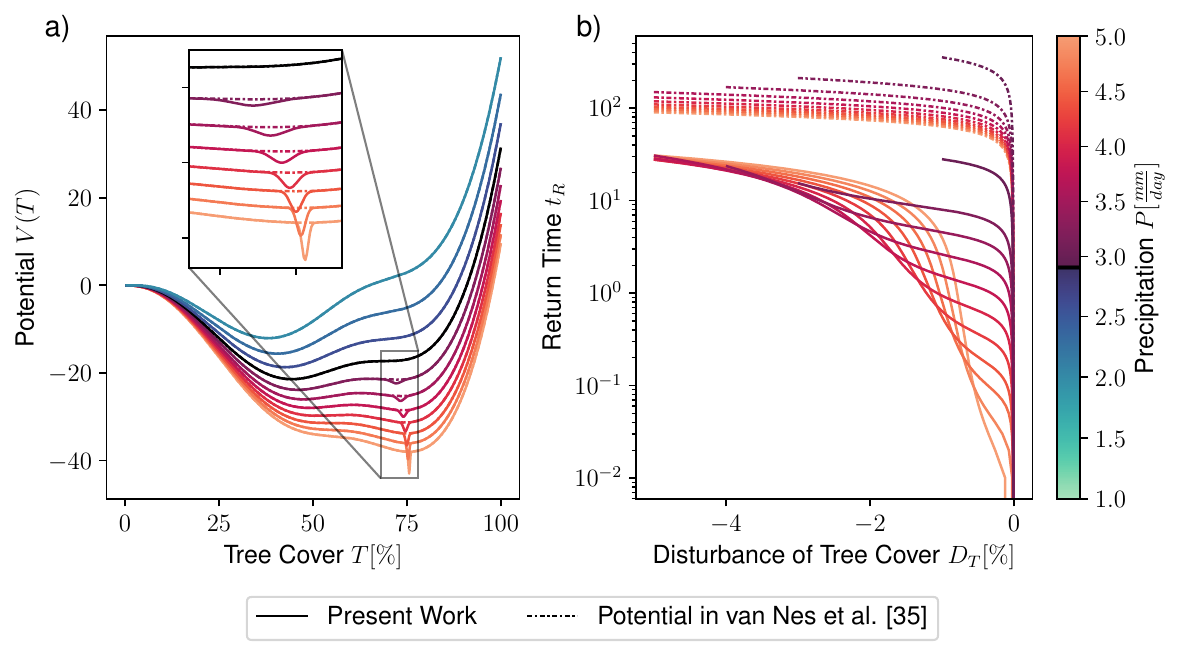} 
   \caption{\textbf{a)} The potential landscape of the tropical vegetation  model used in this study, for varying precipitation levels (colors). The dotted line represents the original potential used in \citep{vannes.etal_2014}. The solid line includes the modifications to represent small time scales introduced in this work. \textbf{b)} The return time it takes for the forest to recover from a disturbance. In the original \citet{vannes.etal_2014} model (dash-dotted), return times even for very small disturbances are in the hundreds of years. Our work adds small time scales (solid), i.e. recovery from small disturbances e.g. due to synoptic-scale weather variations is quick.}
    \label{fig:fig2}
\end{figure}

\section{Methods}\label{sec:methods}

\textbf{Conceptual model of a tropical vegetation ecosystem for simulating regime shifts.}
In this work, we investigate tropical rainforest resilience under power-law noise. Forest resilience is frequently measured with indicators exploiting critical slowing down (CSD), such as the autocorrelation. The phenomenon describes a decreased stability of a system up to the point where the system undergoes a critical transition and switches into another state. Hence, the estimation of forest resilience is typically an identical process as the study of early warning to critical transition, such as early warning of tipping elements in the Earth system.

In order to test the reliability of early warning signals (forest resilience indicators), we simulate rainforest-savannah transitions of a tropical rainforest using the conceptual model by \citet{vannes.etal_2014}. Thus, our theoretical approach provides evidence necessary to understand actual measurements that may be influenced by non-Gaussian noise.

In this model, the tree cover $T$ [\%] of the rainforest is a tri-stable system with forest, savannah and treeless states depending on the precipitation $P$ [mm day$^{-1}$] as an external forcing. Ignoring the treeless state, the model dynamics locally exhibit the following characteristics: for $2$ mm day$^{-1}< P < 2.94$ mm day$^{-1}$ only a stable savannah state exists, for $2.95$ mm day$^{-1} < P < 4.41$ mm day$^{-1}$ both a stable savannah and a stable forest state exist and for $4.42$ mm day$^{-1} < P < 5$ mm day$^{-1}$ only a stable forest state exists (Fig.~\ref{fig:fig1}a). These regimes result from a governing equation \cite{vannes.etal_2014}, which without displaying units reads
\begin{equation}\label{eq:fullamazon}
    \frac{dT}{dt} = \frac{0.3P}{0.5+P} T \left(1-\frac{T}{90}\right) - \frac{1.5T}{T+10} - \frac{0.11\cdot T}{(\frac{T}{64})^7+1} 
\end{equation}
Here, the first term describes a logistic growth of tree cover with a precipitation-dependent expansion coefficient. The second term accounts for the Allee effect: if the tree cover is low, new trees have a harder time growing because they have less protective covering from older trees. The third term introduces a wildfire effect: dense forests are subject to higher fire mortality. While this model represents a strongly stylised way to model the dynamics of tropical vegetation, its simplicity is central to our study as it allows for many simulations of critical transitions.\\

\textbf{Model modification to include forest response on small time scales.} 
Originally, the \citet{vannes.etal_2014} tipping model was developed to model critical transitions on long time scales. To investigate early warning for such critical transitions, a system response on short time scales is necessary because all critical slowing down indicators are based on an increasing disturbance recovery time as a tipping point is approached. In the original \citet{vannes.etal_2014} rainforest model, recovery from even very small disturbances would take hundreds of years (Fig.~\ref{fig:fig2}b, dashed lines). This can be loosely understood as the removal of a few trees and subsequent regrowth. However, small changes in the rainforest state $T$ could also be understood as a response to climate variations, i.e. small fluctuations due to weather. For instance, during a drier than usual period, trees might develop fewer leaves, but already a couple of months later they could completely recover (compare \cite{martinez-ramos.etal_1988, vicente-serrano.etal_2013, linscheid.etal_2020a, sierra.etal_2021}). Note that the forest response to weather can be highly nonlinear, as trees have several regulating mechanisms (e.g. stomatal control or hydraulic resistance), which may render the forest fairly stable even under strong climatic fluctuations. Still, it is reasonable to argue that there is the possibility for the intact rainforest to quickly recover from small perturbations which may loosely be understood as climate variations.

We introduce this net response to small fluctuations into the model by adding an additional short-term resilience term to Eq.~\ref{eq:fullamazon}. This leads to the following changes in the potential: around the stable fixed point, we create an additional potential valley, decreasing in depth as the rainforest-savannah tipping point is approached. Fig.~\ref{fig:fig2}a displays the changes: a little peak parametrized by a Gaussian exponential function centred at the rainforest attractor (right-most minimum of the potential) is subtracted from the original potential (Fig.~\ref{fig:fig2}a, dashed lines). This additional stabilizing force strongly reduces the return times from small disturbances (Fig.~\ref{fig:fig2}b, solid lines). The updated model equations, again displayed without units, are:
\begin{eqnarray}\label{eq:model}
    \frac{dT}{dt} = &\frac{0.3P}{0.5+P} T \left(1-\frac{T}{90}\right) - \frac{1.5T}{T+10} - \frac{0.11\cdot T}{(\frac{T}{64})^7+1} \\
    &- (\tanh(25P - 75) + 1) \frac{5}{2} \frac{\Delta T_{fix}(T,P)}{\exp(5-P)^3}\exp(-\frac{5}{2}\frac{\Delta T_{fix}(T,P)^2}{\exp(5-P)^2})\\
    &+ dN_t
\end{eqnarray}
Here, $\Delta T_{fix}(T, P) = T - T_{fix}(P)$ is the distance to the rainforest fixed point at given $P$. We chose the height and width of the Gaussian such that the return time of a $1\%$ tree cover disturbance at $P=5$  mm day$^{-1}$ is approximately one month. The width is decreased exponentially ($\exp(5-P)$) with lower $P$ and the height is decreased by a scaled sigmoid ($\tanh(25P-75) + 1$) such that close to the savannah transition it vanishes. In addition, the new term $dN_t$ represents noise increments.\\

\textbf{Simulating rainforest disturbances with $\alpha$-stable Lévy noise}\label{sec:levy}
Multiple stressors influence tropical rainforests. Many of these occur as single, discrete, events. For instance, wildfires or windthrows may destroy parts of the forest within a few hours. The outcome is subsequently visible in the canopy gap structure. Commonly, the observed fragmentation patterns of tropical rainforests, in particular the Amazon, follows a power-law-like distribution\cite{fisher.etal_2008,taubert.etal_2018,gloor.etal_2009,negron-juarez.etal_2010,espirito-santo.etal_2010,reis.etal_2022,espirito-santo.etal_2014,chambers.etal_2013,asner.etal_2013,farrior.etal_2016,negron-juarez.etal_2018}. This means, patches of all sizes exist, and particularly large gaps are observed more frequently than they would be if their origin was a random Gaussian noise process. In reality, the disturbance distribution alone does not cause the observed gap structure. Instead, complex spatial and temporal mechanisms are at play. However, as droughts \cite{stocker.etal_2019} and wildfires \cite{nicoletti.etal_2023} can also follow power-law-like distributions, it is possible that random perturbations in tropical rainforests follow a non-Gaussian distribution, possibly with power-law tails.

In this work, we assess via simulations whether critical slowing down can be detected if the underlying noise distribution is not Gaussian but has power-law tails with jumps. For random variables with power-law tails (and thus infinite variance), a generalized central limit theorem holds \cite[in §35~Theorem~5]{gnedenko.kolmogorov_1954}. The limit distributions belong to the family of $\alpha$-stable Lévy noise \cite{levy_1924}. This is a family of distributions with parameters $\alpha \in (0, 2], \beta \in [-1,1]$ that generalizes the Gaussian distribution (containing it at $\alpha = 2, \beta = 0$). We choose $\beta = -1$, to simulate negative disturbances, the resulting tail behaviour in a log-log plot is shown in Fig.~\ref{fig:fig1}b. For $\alpha < 2$, the tails follow a power-law \cite{chechkin.etal_2008,janicki.weron_1994}. In this work, we use $\alpha$-stable Lévy noise $L^\alpha(\sigma; \beta)$ with amplitude $\sigma$ as noise increments in our models:
\begin{equation}
    dN_t \sim L^\alpha(dt^{\frac{1}{\alpha}}\sigma; \beta = -1)
\end{equation}
Such Lévy noise leads to jumps in the forest trajectory if $\alpha < 1$, which could represent single, rapid events such as wildfires \cite{nicoletti.etal_2023} or windthrows \cite{negron-juarez.etal_2018}.\\

\textbf{Early warning of regime shift with critical slowing down indicators}
When a system approaches a tipping point, critical slowing down measures the gradually declining recovery rate of the linearized dynamics and can be used as an early warning indicator \cite{wissel_1984, vannes.scheffer_2007, dakos.etal_2008, scheffer.etal_2009}. Here, critical slowing down refers to an increase in the recovery time from perturbations as the system approaches a bifurcation. Measured by the rate of recovery from small perturbations, the phenomenon is used to assess ecological resilience and to warn before a critical transition is reached. The theoretical justification for using the standard deviation and the autocorrelation as critical slowing down indicators arises from first linearizing the dynamics around a given stable fixed point $x^*$ to obtain the linear Langevin equation

\begin{equation}\label{eq:langevinB}
    dx = \lambda (x - x^*)dt + \sigma dB_t\;
\end{equation}
where $B_t$ is a Wiener process. 
The solution is an Ornstein-Uhlenbeck process, for which analytic expressions of both classical CSD indicators can be derived \cite{boers_2021}:

\begin{eqnarray}
    Var[x] &= -\frac{\sigma^2}{2 \lambda}\label{eq:VarB} \\
    AC1[x] &= \exp(\lambda \Delta t)\label{eq:AC1B},
\end{eqnarray}
where $\Delta t>0$ is the sampling time step.
Clearly, in approach to a critical point at which the linear restoring rate vanishes ($\lambda \nearrow 0$), both indicators increase monotonically ($Var[x] \nearrow \infty$, $AC1[x] \nearrow 1$). This statement holds also for Eq.~\ref{eq:model} with a more general Gaussian noise term $dN_t$. However, if the noise term follows an $\alpha$-stable Lévy distribution with $\alpha<2$, variance and AC(1) are ill-defined because the second moment of $x$ would be infinite. Since the EWS assessments inherent to this study are always performed on a bounded state space, this poses no practical problem and an analogous CSD characteristic to that seen for the Gaussian white noise case can indeed be analytically motivated (see \ref{app:alphaEWS} for further discussion).
We will assess the numerical behaviour of the above indicators using simulations with disturbances that follow such power-law noise with jumps. Furthermore, we assess the performance of the interquartile range (IQR), an indicator of the width of a distribution similar to the standard deviation, yet more robust to outliers. Even though we do not give exact analytical expressions for the IQR, critical slowing down suggests that as a critical point is approached, the IQR increases monotonically ($IQR \nearrow \infty$).

The concrete protocol follows \citet{boers_2021} and is depicted in Fig.~\ref{fig:fig1}c-d. We simulate 1000 model years of steady state, followed by 500 years of climate change and 500 years in a new steady state. During the simulated climate change, the forcing parameter precipitation is changed from $5.0$ mm day$^{-1}$ to $2.5$ mm day$^{-1}$. Hence, the forest undergoes a critical transition with the rainforest state ceasing to exist around 1400 years at a value of $2.943$ mm day$^{-1}$. We simulate realizations of systems under such forcing with sampled noise. To study the actual noise independent of the underlying drift, we subtract a deterministic trajectory from each stochastic realization. Note that this optimal way of nonlinearly detrending the time series -- a necessary processing step prior to computing CSD indicators -- is only possible in model systems. When working with observational data, one has to rely on suitable low-pass filtering. We then compute moving-window early warning signals with a window size of $10$ model years. From these indicators, we compute a slope. We assess if the slope is increasing by comparing it with a null distribution of slopes of the same time series perturbed under random phase shifts. Early warning is then predicted if the observed slope is among the 5\% highest slopes derived from the surrogate time series.

\begin{figure}[t]
    \centering
    \includegraphics[width=\linewidth]{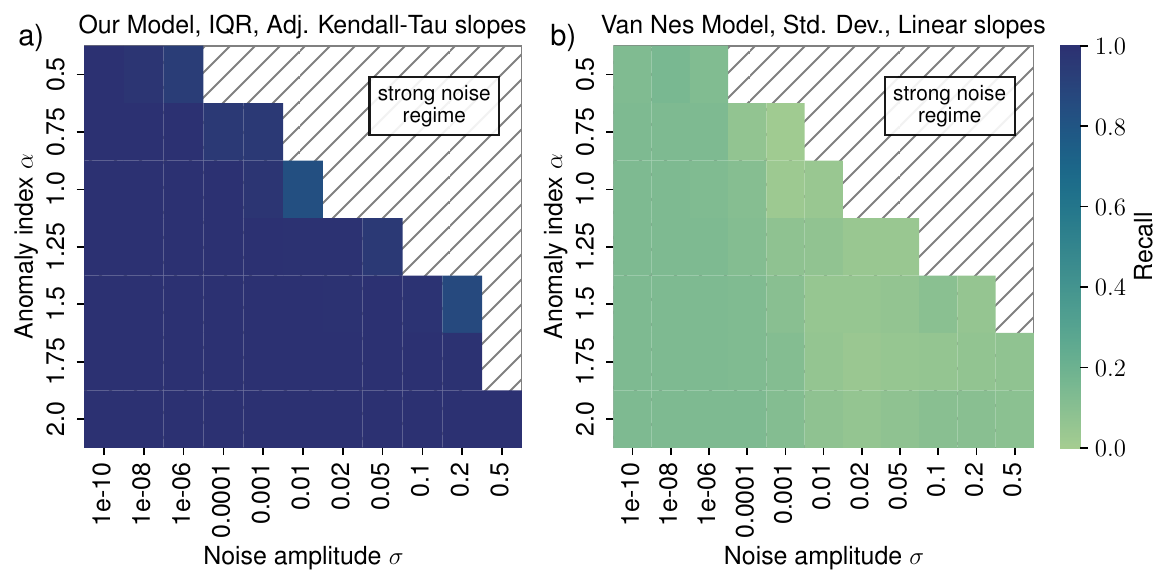} 
   \caption{Recall across various noise amplitudes and anomaly indices ($\alpha = 2$ is Gaussian noise). Shown are only those noise configurations that can be considered as weak noise determined by the observed first passage time. Panel \textbf{a)} shows a processing chain that leads to high recall, i.e. the power-law disturbances are properly dealt with. Recall is close to $1.0$ for all noise configurations, with slight decrease close to the strong noise regime. Panel \textbf{b)} in contrast shows no proper treatment of power-law disturbances, which leads to low recall close to $10\% \pm 2\%$.}
    \label{fig:fig3}
\end{figure}

\begin{table}[t]
    \centering
    \begin{tabularx}{0.75\textwidth}{Xllc@{\hspace{3em}}cc}
    \toprule
    Model & Slope & Indicator & & Recall & FPR \\
    \midrule
    van Nes & Linear & Std. Dev. & & 0.10 & 0.10 \\
    van Nes & Kendall-Tau & IQR & & 0.11 & 0.08 \\
    \midrule
    Ours & Linear & IQR & & 0.85 & 0.10 \\
     Ours & Theil-Sen & IQR & & 0.51 & 0.08 \\
     Ours & Kendall-Tau & IQR & & \textbf{0.99} & 0.10 \\
     \midrule
     Ours & Kendall-Tau & Std. Dev. & & 0.97 & 0.10 \\
     Ours & Kendall-Tau & Lag-1 AC & & 0.96 & 0.11 \\
      \midrule
      Ours & Adj. Kendall-Tau & IQR & & \textbf{0.99} & \textbf{0.02}\\
      \bottomrule
    \end{tabularx}
    \caption{Average recall and false positive rate (FPR) for various processing chains. Averages are computed across a wide choice of noise amplitudes and anomaly indices, as long as the resulting noise can be considered weak (as measured by the first passage time). The first two rows show results for the original \citet{vannes.etal_2014} model, for which critical slowing down does not work due to a lack of representation of small time scales. The second block of rows presents different choices for computing the slope of the early warning indicator when considering our model, which includes small time scales: the non-linear parametric Kendall-Tau slope is robust and thus preferrential. The robustness becomes clear in the third block, there is little variation in recall when comparing other early warning indicators than the interquartile-range (IQR). The final row shows that adjusting the early warning based on the absolute autocorrelation reduces the FPR by a factor of five, while keeping a high recall.}
    \label{tab:tab1}
\end{table}

\begin{figure}[t]
    \centering
    \includegraphics[width=\linewidth]{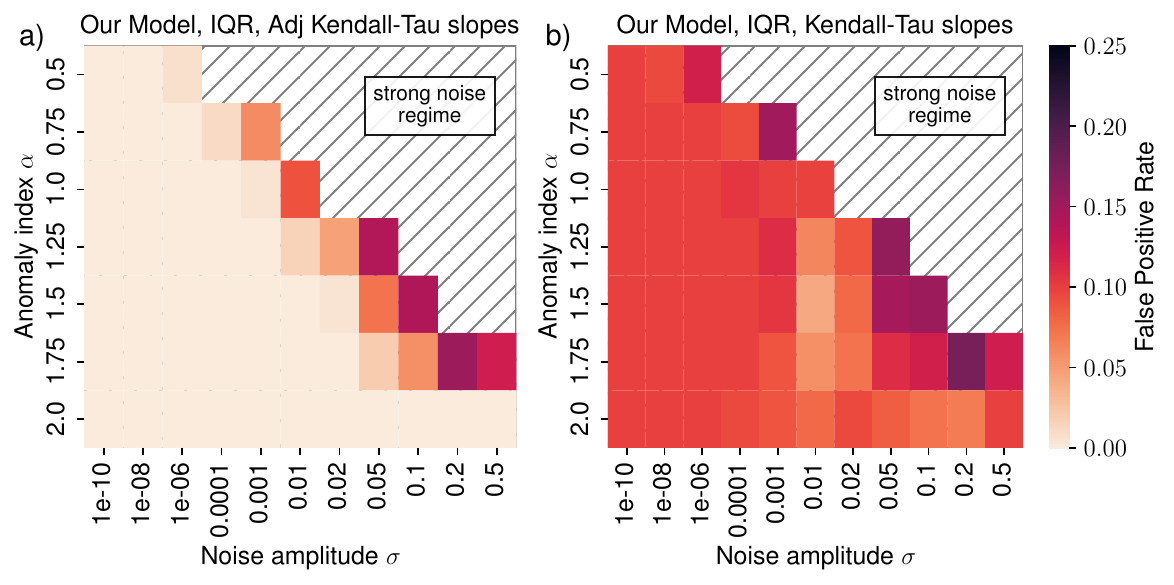} 
   \caption{False positive rate (FPR) for parameter ranges also shown in Fig.~\ref{fig:fig3}. Panel \textbf{a)} displays a processing chain that leads to low FPR, i.e. the power-law disturbances are properly dealt with. FPR is close to $0.0$ for noise configurations far from the strong noise regime, but increases close to the strong noise regime. Panel \textbf{b)} in contrast shows the effect of improper treatment of power-law disturbances. This leads to a false positive rate close to $10\%$, but with a higher variation across noise configurations.}
    \label{fig:fig4}
\end{figure}

\section{Results}
Early warning with few misses and few false alarms is possible under Lévy noise according to our experiments. We quantify the rate of achieving accurate early warnings with the recall, the fraction of correctly predicted transitions (true positives, $TP$) over all transitions (true positives and false negatives, $FN$). In other terms, the recall is high if no critical transition is missed, and low, if there are too few early warnings.
\begin{equation}
    Recall = \frac{TP}{TP+FN}
\end{equation}
To study false alarms, we assess our early warning classification pipeline over random 50-year windows of the simulated steady-state years (which have no critical transition). We quantify the rate of false alarms with the false positive rate (FPR), i.e. the fraction of transitions predicted (false positives, $FP$) over all assessed transition-free time windows (false positives and true negatives, $TN$). In other terms, the FPR is low if there are no false alarms, and high, if there are too many early warnings.
\begin{equation}
    FPR = \frac{FP}{FP+TN}
\end{equation}

When resolving small timescales and assessing critical slowing down with the non-parametric Kendall-$\tau$ slope, we find an high average recall of $0.99$ (fig.~\ref{fig:fig3}a). Hence, early warning works with few misses. This corresponds to a measure of forest resilience that is able to detect destabilization, i.e. a decline in the system's resilience. More specifically, the early warning pipeline displays a high recall (on average $0.99$) across all combinations of noise amplitude $\sigma$ and Lévy index $\alpha$ that can be considered as weak noise levels. This suggests that within the employed limits both parameters do not strongly affect the performance of the CSD indicators. The depicted early warning pipeline entails the usage of the interquartile range as an early warning indicator and the Kendall-$\tau$ to assess an increasing slope. In addition, the underlying time series data was generated by our model modification which resolves small timescales.

In cases that small timescales are not resolved, we obtain a low average recall of $0.11$ (fig.~\ref{fig:fig3}b). Moreover, the recall is low across all admissible parameter combinations. Thus, also for Gaussian noise ($\alpha = 2$), the early warning does not work, which underlines the necessity to resolve small timescales. In other terms, if this were the case for the real world, i.e. the rainforest would only respond very slowly to perturbations, a declining forest resilience would not be measurable with CSD indicators. For such a transient situation, even considering weak Gaussian noise does not allow us to measure forest resilience solely on the basis of CSD indicators. Hence, in the following, we will only consider modelled time series, where a response to disturbances on small timescales is resolved.

If critical slowing down is assessed on a filtered time series, we find a low false positive rate of on average $0.02$ for the non-parametric Kendall-$\tau$ slope on IQR time series (fig.~\ref{fig:fig4}a). Here, the filtering is leveraged as a post-hoc adjustment to assess the significance of a decline: by random chance, one might detect a significant slope even if the magnitude of the autocorrelation is very low. Hence, we flag those early warnings, where the temporal autocorrelation is below $0.5$. We refer to the resulting indicator as \emph{Adjusted Kendall-Tau Slope}. Note that the autocorrelation is particularly suitable to perform this filtering because there is a natural way of thresholding, in contrast to the standard deviation or the IQR. When including such filtering, the early warning (forest resilience) pipeline produces few false alarms.

In constrast, when early warning is detected on raw (unfiltered) time series, we detect a non-neglibile average false positive rate of $0.10$, again for the case of non-parametric Kendall-$\tau$ slope on IQR time series  (fig.~\ref{fig:fig4}b). Hence, the early warning pipeline without the filtering of perturbations according to their temporal autocorrelation, raises false alarms. For an ecological system, this means a declining resilience measured solely by a CSD indicator should always be double-checked, for instance against the magnitude of the temporal autocorrelation in historical time series. Otherwise, a false alarm may be raised. 

Table~\ref{tab:tab1} summarizes the key findings by comparing mean recall and mean false positive rates across a selection of scenarios. First, using Kendall-tau slopes in our model (including small time scales), high recall can be achieved (sixth row). Second, adjusting the early warning for the magnitude of the autocorrelation, low false-positive rates at high recall are possible (last row). For the original \citet{vannes.etal_2014} model, recall is always low, irrespective of the processing (exemplified in the first two rows, but consistent among other scenarios, not displayed in Table~\ref{tab:tab1}). For our model, only the Kendall-tau slope gives consistently high recall, linear and Theil-Sen slopes suffer from jumps in the noise structure (second block of rows). The type of indicator used to assess CSD (i.e. rainforest resilience) is less relevant, we find the interquartile range (IQR) slightly outperforms standard deviation and autocorrelation, but all three measures are valid options (third block of rows). The average false positive rate before adjusting is around $10\%$, and can be reduced to only $2\%$ by removing false positives with low autocorrelation.

\section{Discussion}
Our work suggests that critical slowing down can be used to assess resilience loss due to an approaching critical transition also in the presence of extreme events modelled by $\alpha$-stable Lévy noise. In the case of the Amazon rainforest, this confirms the empirical results by other studies indicating a destabilization of the system \cite{boulton.etal_2022,nobre.etal_2016,lovejoy.nobre_2018,lovejoy.nobre_2019,flores.holmgren_2021} as being more generally applicable than previously thought. 

The particular tropical rainforest model used in this study is a strongly simplified rainforest model. However, it can be understood as a special case of a very general model: a double-well potential, the mathematical normal form of a saddle-node bifurcation. For transitions in a double-well potential, our findings hold and are robust (see \ref{sec:dw}). Hence, they are not limited to just the particular choice of the conceptual model in this study but are more widely applicable to any system exhibiting such critical transitions. For instance, many tipping elements of the Earth system have been modelled through special cases of the double-well potential \cite{lenton.etal_2008,wunderling.etal_2021, wunderling.etal_2023a}. Also some other models of tropical rainforests \cite{staal.etal_2015} belong to the double-well potential family. Furthermore, we claim our analysis is relevant for the actual rainforest system, which is highly complex. To support such statement, one may consider increasing complexity of the model, e.g. by studying global dynamical vegetation models. Due to their high-dimensionality, they usually cannot be easily represented by a potential landscape. However, for CMIP6 models, abrupt local forest dieback has been diagnosed in the Amazon basin \cite{parry.etal_2022}. Hence, the existence of critical transitions is indicated and locally such a transition can again be reasonably well approximated by simple double-well potential models, as done in this study.

Not all non-Gaussian disturbances necessarily are of $\alpha$-stable Lévy-type. Instead, in some circumstances, coloured noise has been observed in ecological systems and in particular in tropical rainforest \cite{halley_1996, halley.inchausti_2004, newbery.etal_2011, vasseur.yodzis_2004, yang.etal_2019b}. In such cases, the two most commonly postulated colour noises are pink noise and red noise. For red noise, which displays an auto-correlated noise structure, resilience measures based on CSD indicators need to be adapted, but then they work as in the case of white noise \cite{morr.boers_2023a}. Pink noise has power-law tails but does not lead to jumps in the forest state evolution. Therefore, we find that our results for $\alpha$-stable Lévy noise are robust for pink noise (see \ref{sec:pink} and Table~\ref{tab:tab1pink} for details). Hence, for a wide range of non-Gaussian noise observed in ecology, CSD indicators are valid choices to measure tropical rainforest resilience.

Two further points regarding the accuracy of our mathematical model require caution: First, we have chosen to model forest disturbances via $\alpha$-stable Lévy noise to honour the prevalence of extreme events in observations. However, as the variable representing the state of the rainforest is bounded between $0$ and $100 \%$ (tree cover density), the far end of the noise tail needs to be disregarded, because tree cover cannot fall below $0\%$ nor go beyond $100\%$. This restricts the conceptual modelling capabilities of the $\alpha$-stable noise model. It does however not pose a practical problem for interpreting the model data as observations from a multi-stable forest system because an exceedingly extreme disturbance always simply leads to a tipped system. Our work demonstrates that the concept of CSD holds in the case of $\alpha$-stable noise distributions when observing the time span before tipping. Second, due to ongoing deforestation for many decades, one might argue the Amazon is not in a steady state, but rather on a transient. Hence, the equilibrium assumption of critical slowing down, whereby a slow change in the external forcing only changes the equilibrium state, but does not keep the system in disequilibrium for long, may be violated. Nevertheless, studies have found that the Amazon rainforest may approach a tipping point due to deforestation \cite{boers.etal_2017b, bochow.boers_2023}.

\section{Conclusion}
We find robustness of critical slowing down indicators used to measure resilience in multi-stable systems driven by $\alpha$-stable Lévy noise. The presence of disturbances with power-law spectrum occurring in single discrete jumps does not affect the identification of early warning of a critical transition as long as the jump size is small enough to avoid immediate, noise-induced transitions between alternative stable states. For this purpose, we find it is ideal to test the interquartile range with Kendall-Tau slope for significant increases and filter for cases with low autocorrelation. With such processing, high recall and low false positive rates can be achieved. Most recent work computing resilience indicators based on remote sensing in the Amazon basin follow a similar procedure \cite{boulton.etal_2022,smith.etal_2022}. In particular, they assess rainforest resilience with critical slowing down indicators and find a resilience decline. Our work emphasizes that such empirical findings are not corroborated in the presence of non-Gaussian disturbances, which could not be ruled out previously. Hence, it adds to the increasing evidence that the Amazon rainforest's resilience has been declining in recent decades, irrespective of the actual nature of the noise (white, coloured, or Lévy noise). Future work may extend the analysis to include a spatial component. Spatial processes are highly relevant for tropical vegetation health and resulting spatial patterns \cite{bastiaansen.etal_2020, rietkerk.etal_2021}, e.g. patchiness, have been introduced as another type of early warning signal (i.e. resilience indicator) \cite{dakos.etal_2010}.


\section*{Acknowledgements}
 N.W. and J.F.D. acknowledge support from the European Research Council Advanced Grant project ERA (Earth Resilience in the Anthropocene, ERC-2016-ADG-743080). J.F.D. is grateful for financial support by the  German Federal Ministry for Education and Research (BMBF) (project ‘PIK Change’, grant 01LS2001A). N.B. acknowledges funding by the Volkswagen foundation and by the European Union’s Horizon 2020 research and innovation programme under the Marie Sklodowska-Curie grant agreement No.956170, and under grant agreement No. 820970. A.S. was supported by the Dutch Research Council (NWO) Talent Program Grant VI.Veni.202.170. M.H. thanks the Serrapilheira Institute (grant number Serra-1709-18983).

\section*{Code to reproduce results} \url{https://github.com/vitusbenson/tropical_rainforest_resilience_nongaussian}

\section*{References}
\printbibliography[heading=none]

@article{armstrongmckay.etal_2022,
  title = {Exceeding 1.5°{{C}} Global Warming Could Trigger Multiple Climate Tipping Points},
  author = {Armstrong McKay, David I. and Staal, Arie and Abrams, Jesse F. and Winkelmann, Ricarda and Sakschewski, Boris and Loriani, Sina and Fetzer, Ingo and Cornell, Sarah E. and Rockström, Johan and Lenton, Timothy M.},
  date = {2022-09-09},
  journaltitle = {Science},
  shortjournal = {Science},
  volume = {377},
  number = {6611},
  pages = {eabn7950},
  issn = {0036-8075, 1095-9203},
  doi = {10.1126/science.abn7950},
  url = {https://www.science.org/doi/10.1126/science.abn7950},
  urldate = {2022-11-19},
  langid = {english}
}

@article{asner.etal_2013,
  title = {Forest {{Canopy Gap Distributions}} in the {{Southern Peruvian Amazon}}},
  author = {Asner, Gregory P. and Kellner, James R. and Kennedy-Bowdoin, Ty and Knapp, David E. and Anderson, Christopher and Martin, Roberta E.},
  date = {2013-04-15},
  journaltitle = {PLOS ONE},
  shortjournal = {PLOS ONE},
  volume = {8},
  number = {4},
  pages = {e60875},
  publisher = {{Public Library of Science}},
  issn = {1932-6203},
  doi = {10.1371/journal.pone.0060875},
  url = {https://journals.plos.org/plosone/article?id=10.1371/journal.pone.0060875},
  urldate = {2022-11-30},
  langid = {english}
}

@article{bastiaansen.etal_2020,
  title = {The Effect of Climate Change on the Resilience of Ecosystems with Adaptive Spatial Pattern Formation},
  author = {Bastiaansen, Robbin and Doelman, Arjen and Eppinga, Maarten B. and Rietkerk, Max},
  date = {2020},
  journaltitle = {Ecology Letters},
  volume = {23},
  number = {3},
  pages = {414--429},
  issn = {1461-0248},
  doi = {10.1111/ele.13449},
  url = {https://onlinelibrary.wiley.com/doi/abs/10.1111/ele.13449},
  urldate = {2023-10-12},
  langid = {english}
}

@article{bochow.boers_2023,
  title = {The {{South American}} Monsoon Approaches a Critical Transition in Response to Deforestation},
  author = {Bochow, Nils and Boers, Niklas},
  date = {2023-10-04},
  journaltitle = {Science Advances},
  volume = {9},
  number = {40},
  pages = {eadd9973},
  publisher = {{American Association for the Advancement of Science}},
  doi = {10.1126/sciadv.add9973},
  url = {https://www.science.org/doi/full/10.1126/sciadv.add9973},
  urldate = {2023-10-12}
}

@article{boers_2021,
  title = {Observation-Based Early-Warning Signals for a Collapse of the {{Atlantic Meridional Overturning Circulation}}},
  author = {Boers, Niklas},
  date = {2021-08},
  journaltitle = {Nature Climate Change},
  shortjournal = {Nat. Clim. Chang.},
  volume = {11},
  number = {8},
  pages = {680--688},
  issn = {1758-678X, 1758-6798},
  doi = {10.1038/s41558-021-01097-4},
  url = {https://www.nature.com/articles/s41558-021-01097-4},
  urldate = {2022-01-31},
  langid = {english}
}

@article{boers.etal_2017b,
  title = {A Deforestation-Induced Tipping Point for the {{South American}} Monsoon System},
  author = {Boers, Niklas and Marwan, Norbert and Barbosa, Henrique M. J. and Kurths, Jürgen},
  date = {2017-01-25},
  journaltitle = {Scientific Reports},
  shortjournal = {Sci Rep},
  volume = {7},
  number = {1},
  pages = {41489},
  publisher = {{Nature Publishing Group}},
  issn = {2045-2322},
  doi = {10.1038/srep41489},
  url = {https://www.nature.com/articles/srep41489},
  urldate = {2023-10-12},
  issue = {1},
  langid = {english}
}

@article{boers.etal_2022,
  title = {Theoretical and Paleoclimatic Evidence for Abrupt Transitions in the {{Earth}} System},
  author = {Boers, Niklas and Ghil, Michael and Stocker, Thomas F.},
  date = {2022-09},
  journaltitle = {Environmental Research Letters},
  shortjournal = {Environ. Res. Lett.},
  volume = {17},
  number = {9},
  pages = {093006},
  publisher = {{IOP Publishing}},
  issn = {1748-9326},
  doi = {10.1088/1748-9326/ac8944},
  url = {https://dx.doi.org/10.1088/1748-9326/ac8944},
  urldate = {2022-11-19},
  langid = {english}
}

@article{boettner.boers_2022,
  title = {Critical Slowing down in Dynamical Systems Driven by Nonstationary Correlated Noise},
  author = {Boettner, Christopher and Boers, Niklas},
  date = {2022-03-28},
  journaltitle = {Physical Review Research},
  shortjournal = {Phys. Rev. Research},
  volume = {4},
  number = {1},
  pages = {013230},
  issn = {2643-1564},
  doi = {10.1103/PhysRevResearch.4.013230},
  url = {https://link.aps.org/doi/10.1103/PhysRevResearch.4.013230},
  urldate = {2022-05-18},
  langid = {english}
}

@article{boulton.etal_2022,
  title = {Pronounced Loss of {{Amazon}} Rainforest Resilience since the Early 2000s},
  author = {Boulton, Chris A. and Lenton, Timothy M. and Boers, Niklas},
  date = {2022-03},
  journaltitle = {Nature Climate Change},
  shortjournal = {Nat. Clim. Chang.},
  volume = {12},
  number = {3},
  pages = {271--278},
  publisher = {{Nature Publishing Group}},
  issn = {1758-6798},
  doi = {10.1038/s41558-022-01287-8},
  url = {https://www.nature.com/articles/s41558-022-01287-8},
  urldate = {2022-06-20},
  issue = {3},
  langid = {english}
}

@article{chambers.etal_2013,
  title = {The Steady-State Mosaic of Disturbance and Succession across an Old-Growth {{Central Amazon}} Forest Landscape},
  author = {Chambers, Jeffrey Q. and Negron-Juarez, Robinson I. and Marra, Daniel Magnabosco and Di Vittorio, Alan and Tews, Joerg and Roberts, Dar and Ribeiro, Gabriel H. P. M. and Trumbore, Susan E. and Higuchi, Niro},
  date = {2013-03-05},
  journaltitle = {Proceedings of the National Academy of Sciences},
  volume = {110},
  number = {10},
  pages = {3949--3954},
  publisher = {{Proceedings of the National Academy of Sciences}},
  doi = {10.1073/pnas.1202894110},
  url = {https://www.pnas.org/doi/abs/10.1073/pnas.1202894110},
  urldate = {2022-11-30}
}

@incollection{chechkin.etal_2008,
  title = {Introduction to the {{Theory}} of {{Lévy Flights}}},
  booktitle = {Anomalous {{Transport}}},
  author = {Chechkin, Alexei V. and Metzler, Ralf and Klafter, Joseph and Gonchar, Vsevolod Yu.},
  editor = {Klages, Rainer and Radons, Günter and Sokolov, Igor M.},
  date = {2008-07-23},
  pages = {129--162},
  publisher = {{Wiley-VCH Verlag GmbH \& Co. KGaA}},
  location = {{Weinheim, Germany}},
  doi = {10.1002/9783527622979.ch5},
  url = {https://onlinelibrary.wiley.com/doi/10.1002/9783527622979.ch5},
  urldate = {2022-05-10},
  isbn = {978-3-527-62297-9 978-3-527-40722-4},
  langid = {english}
}

@article{ciemer.etal_2019,
  title = {Higher Resilience to Climatic Disturbances in Tropical Vegetation Exposed to More Variable Rainfall},
  author = {Ciemer, Catrin and Boers, Niklas and Hirota, Marina and Kurths, Jürgen and Müller-Hansen, Finn and Oliveira, Rafael S. and Winkelmann, Ricarda},
  date = {2019-03},
  journaltitle = {Nature Geoscience},
  shortjournal = {Nat. Geosci.},
  volume = {12},
  number = {3},
  pages = {174--179},
  publisher = {{Nature Publishing Group}},
  issn = {1752-0908},
  doi = {10.1038/s41561-019-0312-z},
  url = {https://www.nature.com/articles/s41561-019-0312-z},
  urldate = {2023-02-09},
  issue = {3},
  langid = {english}
}

@article{dakos.etal_2008,
  title = {Slowing down as an Early Warning Signal for Abrupt Climate Change},
  author = {Dakos, Vasilis and Scheffer, Marten and family=Nes, given=Egbert H, prefix=van, useprefix=true and Brovkin, Victor and Petoukhov, Vladimir and Held, Hermann},
  date = {2008},
  journaltitle = {PNAS},
  volume = {105},
  number = {38},
  pages = {5},
  doi = {10.1073/pnas.0802430105},
  langid = {english}
}

@article{dakos.etal_2010,
  title = {Spatial Correlation as Leading Indicator of Catastrophic Shifts},
  author = {Dakos, Vasilis and family=Nes, given=Egbert H., prefix=van, useprefix=true and Donangelo, Raúl and Fort, Hugo and Scheffer, Marten},
  date = {2010-08},
  journaltitle = {Theoretical Ecology},
  shortjournal = {Theor Ecol},
  volume = {3},
  number = {3},
  pages = {163--174},
  issn = {1874-1738, 1874-1746},
  doi = {10.1007/s12080-009-0060-6},
  url = {http://link.springer.com/10.1007/s12080-009-0060-6},
  urldate = {2022-01-31},
  langid = {english}
}

@article{davidson.etal_2012,
  title = {The {{Amazon}} Basin in Transition},
  author = {Davidson, Eric A. and family=Araújo, given=Alessandro C., prefix=de, useprefix=true and Artaxo, Paulo and Balch, Jennifer K. and Brown, I. Foster and C. Bustamante, Mercedes M. and Coe, Michael T. and DeFries, Ruth S. and Keller, Michael and Longo, Marcos and Munger, J. William and Schroeder, Wilfrid and Soares-Filho, Britaldo S. and Souza, Carlos M. and Wofsy, Steven C.},
  date = {2012-01},
  journaltitle = {Nature},
  volume = {481},
  number = {7381},
  pages = {321--328},
  publisher = {{Nature Publishing Group}},
  issn = {1476-4687},
  doi = {10.1038/nature10717},
  url = {https://www.nature.com/articles/nature10717},
  urldate = {2022-11-19},
  issue = {7381},
  langid = {english}
}

@article{dutta.etal_2018,
  title = {Robustness of Early Warning Signals for Catastrophic and Non-Catastrophic Transitions},
  author = {Dutta, Partha Sharathi and Sharma, Yogita and Abbott, Karen C.},
  date = {2018},
  journaltitle = {Oikos},
  volume = {127},
  number = {9},
  pages = {1251--1263},
  issn = {1600-0706},
  doi = {10.1111/oik.05172},
  url = {https://onlinelibrary.wiley.com/doi/abs/10.1111/oik.05172},
  urldate = {2023-09-08},
  langid = {english}
}

@article{espirito-santo.etal_2010,
  title = {Storm Intensity and Old-Growth Forest Disturbances in the {{Amazon}} Region},
  author = {Espírito-Santo, F. D. B. and Keller, M. and Braswell, B. and Nelson, B. W. and Frolking, S. and Vicente, G.},
  date = {2010},
  journaltitle = {Geophysical Research Letters},
  volume = {37},
  number = {11},
  issn = {1944-8007},
  doi = {10.1029/2010GL043146},
  url = {https://onlinelibrary.wiley.com/doi/abs/10.1029/2010GL043146},
  urldate = {2022-11-30},
  langid = {english}
}

@article{espirito-santo.etal_2014,
  title = {Size and Frequency of Natural Forest Disturbances and the {{Amazon}} Forest Carbon Balance},
  author = {Espírito-Santo, Fernando D. B. and Gloor, Manuel and Keller, Michael and Malhi, Yadvinder and Saatchi, Sassan and Nelson, Bruce and Junior, Raimundo C. Oliveira and Pereira, Cleuton and Lloyd, Jon and Frolking, Steve and Palace, Michael and Shimabukuro, Yosio E. and Duarte, Valdete and Mendoza, Abel Monteagudo and López-González, Gabriela and Baker, Tim R. and Feldpausch, Ted R. and Brienen, Roel J. W. and Asner, Gregory P. and Boyd, Doreen S. and Phillips, Oliver L.},
  date = {2014-03-18},
  journaltitle = {Nature Communications},
  shortjournal = {Nat Commun},
  volume = {5},
  number = {1},
  pages = {3434},
  publisher = {{Nature Publishing Group}},
  issn = {2041-1723},
  doi = {10.1038/ncomms4434},
  url = {https://www.nature.com/articles/ncomms4434;},
  urldate = {2022-11-30},
  issue = {1},
  langid = {english}
}

@article{farrior.etal_2016,
  title = {Dominance of the Suppressed: {{Power-law}} Size Structure in Tropical Forests},
  shorttitle = {Dominance of the Suppressed},
  author = {Farrior, C. E. and Bohlman, S. A. and Hubbell, S. and Pacala, S. W.},
  date = {2016-01-08},
  journaltitle = {Science},
  volume = {351},
  number = {6269},
  pages = {155--157},
  publisher = {{American Association for the Advancement of Science}},
  doi = {10.1126/science.aad0592},
  url = {https://www.science.org/doi/10.1126/science.aad0592},
  urldate = {2022-11-30}
}

@article{fisher.etal_2008,
  title = {Clustered Disturbances Lead to Bias in Large-Scale Estimates Based on Forest Sample Plots},
  author = {Fisher, Jeremy I. and Hurtt, George C. and Thomas, R. Quinn and Chambers, Jeffrey Q.},
  date = {2008},
  journaltitle = {Ecology Letters},
  volume = {11},
  number = {6},
  pages = {554--563},
  issn = {1461-0248},
  doi = {10.1111/j.1461-0248.2008.01169.x},
  url = {https://onlinelibrary.wiley.com/doi/abs/10.1111/j.1461-0248.2008.01169.x},
  urldate = {2022-07-13},
  langid = {english}
}

@article{flores.holmgren_2021,
  title = {White-{{Sand Savannas Expand}} at the {{Core}} of the {{Amazon After Forest Wildfires}}},
  author = {Flores, Bernardo M. and Holmgren, Milena},
  date = {2021-11-01},
  journaltitle = {Ecosystems},
  shortjournal = {Ecosystems},
  volume = {24},
  number = {7},
  pages = {1624--1637},
  issn = {1435-0629},
  doi = {10.1007/s10021-021-00607-x},
  url = {https://doi.org/10.1007/s10021-021-00607-x},
  urldate = {2023-07-18},
  langid = {english}
}

@article{forzieri.etal_2022,
  title = {Emerging Signals of Declining Forest Resilience under Climate Change},
  author = {Forzieri, Giovanni and Dakos, Vasilis and McDowell, Nate G. and Ramdane, Alkama and Cescatti, Alessandro},
  date = {2022-08},
  journaltitle = {Nature},
  volume = {608},
  number = {7923},
  pages = {534--539},
  publisher = {{Nature Publishing Group}},
  issn = {1476-4687},
  doi = {10.1038/s41586-022-04959-9},
  url = {https://www.nature.com/articles/s41586-022-04959-9},
  urldate = {2023-01-10},
  issue = {7923},
  langid = {english}
}

@article{gloor.etal_2009,
  title = {Does the Disturbance Hypothesis Explain the Biomass Increase in Basin-Wide {{Amazon}} Forest Plot Data?},
  author = {Gloor, M. and Phillips, O. L. and Lloyd, J. J. and Lewis, S. L. and Malhi, Y. and Baker, T. R. and López-Gonzalez, G. and Peacock, J. and Almeida, S. and De Oliveira, A. C. Alves and Alvarez, E. and Amaral, I. and Arroyo, L. and Aymard, G. and Banki, O. and Blanc, L. and Bonal, D. and Brando, P. and Chao, K.-J. and Chave, J. and Dávila, N. and Erwin, T. and Silva, J. and Di Fiore, A. and Feldpausch, T. R. and Freitas, A. and Herrera, R. and Higuchi, N. and Honorio, E. and Jiménez, E. and Killeen, T. and Laurance, W. and Mendoza, C. and Monteagudo, A. and Andrade, A. and Neill, D. and Nepstad, D. and Vargas, P. Núñez and Peñuela, M. C. and Cruz, A. Peña and Prieto, A. and Pitman, N. and Quesada, C. and Salomão, R. and Silveira, Marcos and Schwarz, M. and Stropp, J. and Ramírez, F. and Ramírez, H. and Rudas, A. and Ter Steege, H. and Silva, N. and Torres, A. and Terborgh, J. and Vásquez, R. and Van Der Heijden, G.},
  date = {2009},
  journaltitle = {Global Change Biology},
  volume = {15},
  number = {10},
  pages = {2418--2430},
  issn = {1365-2486},
  doi = {10.1111/j.1365-2486.2009.01891.x},
  url = {https://onlinelibrary.wiley.com/doi/abs/10.1111/j.1365-2486.2009.01891.x},
  urldate = {2022-11-30},
  langid = {english}
}

@book{gnedenko.kolmogorov_1954,
  title = {Limit Distributions for Sums of Independent Random Variables},
  author = {Gnedenko, B. V. and Kolmogorov, A. N.},
  date = {1954},
  publisher = {{Addison-Wesley Pub. Co.}},
  location = {{Cambridge, Massachussets}}
}

@article{halley_1996,
  title = {Ecology, Evolution and 1/f -Noise},
  author = {Halley, John M.},
  date = {1996-01-01},
  journaltitle = {Trends in Ecology \& Evolution},
  shortjournal = {Trends in Ecology \& Evolution},
  volume = {11},
  number = {1},
  eprint = {21237757},
  eprinttype = {pmid},
  pages = {33--37},
  publisher = {{Elsevier}},
  issn = {0169-5347},
  doi = {10.1016/0169-5347(96)81067-6},
  url = {https://www.cell.com/trends/ecology-evolution/abstract/0169-5347(96)81067-6},
  urldate = {2023-09-30},
  langid = {english}
}

@article{halley.inchausti_2004,
  title = {The Increasing Importance of 1/f-Noises as Models of Ecological Variability},
  author = {Halley, John M. and Inchausti, Pablo},
  date = {2004-06},
  journaltitle = {Fluctuation and Noise Letters},
  shortjournal = {Fluct. Noise Lett.},
  volume = {04},
  number = {02},
  pages = {R1-R26},
  publisher = {{World Scientific Publishing Co.}},
  issn = {0219-4775},
  doi = {10.1142/S0219477504001884},
  url = {https://www.worldscientific.com/doi/abs/10.1142/s0219477504001884},
  urldate = {2023-09-30}
}

@article{hirota.etal_2011,
  title = {Global {{Resilience}} of {{Tropical Forest}} and {{Savanna}} to {{Critical Transitions}}},
  author = {Hirota, Marina and Holmgren, Milena and Van Nes, Egbert H. and Scheffer, Marten},
  date = {2011-10-14},
  journaltitle = {Science},
  volume = {334},
  number = {6053},
  pages = {232--235},
  publisher = {{American Association for the Advancement of Science}},
  doi = {10.1126/science.1210657},
  url = {https://www.science.org/doi/full/10.1126/science.1210657},
  urldate = {2022-11-20}
}

@book{janicki.weron_1994,
  title = {Simulation and Chaotic Behavior of [Alpha]-Stable Stochastic Processes},
  author = {Janicki, Aleksander and Weron, A.},
  date = {1994},
  series = {Monographs and Textbooks in Pure and Applied Mathematics},
  number = {178},
  publisher = {{M. Dekker}},
  location = {{New York}},
  isbn = {978-0-8247-8882-7},
  langid = {english},
  pagetotal = {355}
}

@article{krakovska.etal_2023,
  title = {Resilience of Dynamical Systems},
  author = {Krakovská, Hana and Kuehn, Christian and Longo, Iacopo P.},
  date = {2023-06-05},
  journaltitle = {European Journal of Applied Mathematics},
  pages = {1--46},
  publisher = {{Cambridge University Press}},
  issn = {0956-7925, 1469-4425},
  doi = {10.1017/S0956792523000141},
  url = {https://www.cambridge.org/core/journals/european-journal-of-applied-mathematics/article/resilience-of-dynamical-systems/B277FB38B049FD4DECC2097E7460E4E3?utm_campaign=shareaholic&utm_medium=twitter&utm_source=socialnetwork#authors-details},
  urldate = {2023-06-16},
  langid = {english}
}

@article{lenton.etal_2008,
  title = {Tipping Elements in the {{Earth}}'s Climate System},
  author = {Lenton, T. M. and Held, H. and Kriegler, E. and Hall, J. W. and Lucht, W. and Rahmstorf, S. and Schellnhuber, H. J.},
  date = {2008-02-12},
  journaltitle = {Proceedings of the National Academy of Sciences},
  shortjournal = {Proceedings of the National Academy of Sciences},
  volume = {105},
  number = {6},
  pages = {1786--1793},
  issn = {0027-8424, 1091-6490},
  doi = {10.1073/pnas.0705414105},
  url = {http://www.pnas.org/cgi/doi/10.1073/pnas.0705414105},
  urldate = {2022-01-31},
  langid = {english}
}

@article{lenton.etal_2022,
  title = {A Resilience Sensing System for the Biosphere},
  author = {Lenton, Timothy M. and Buxton, Joshua E. and Armstrong McKay, David I. and Abrams, Jesse F. and Boulton, Chris A. and Lees, Kirsten and Powell, Thomas W. R. and Boers, Niklas and Cunliffe, Andrew M. and Dakos, Vasilis},
  date = {2022-06-27},
  journaltitle = {Philosophical Transactions of the Royal Society B: Biological Sciences},
  volume = {377},
  number = {1857},
  pages = {20210383},
  publisher = {{Royal Society}},
  doi = {10.1098/rstb.2021.0383},
  url = {https://royalsocietypublishing.org/doi/full/10.1098/rstb.2021.0383},
  urldate = {2023-04-25}
}

@article{levy_1924,
  title = {Théorie des erreurs. La loi de Gauss et les lois exceptionnelles},
  author = {Lévy, P.},
  date = {1924},
  journaltitle = {Bulletin de la Societe Matematique de France},
  shortjournal = {Bul. Soc. Math. France},
  volume = {2},
  pages = {49--85},
  issn = {0037-9484, 2102-622X},
  doi = {10.24033/bsmf.1046},
  url = {http://www.numdam.org/item?id=BSMF_1924__52__49_1},
  urldate = {2022-06-12},
  langid = {french}
}

@article{linscheid.etal_2020a,
  title = {Towards a Global Understanding of Vegetation–Climate Dynamics at Multiple Timescales},
  author = {Linscheid, Nora and Estupinan-Suarez, Lina M. and Brenning, Alexander and Carvalhais, Nuno and Cremer, Felix and Gans, Fabian and Rammig, Anja and Reichstein, Markus and Sierra, Carlos A. and Mahecha, Miguel D.},
  date = {2020-02-24},
  journaltitle = {Biogeosciences},
  volume = {17},
  number = {4},
  pages = {945--962},
  publisher = {{Copernicus GmbH}},
  issn = {1726-4170},
  doi = {10.5194/bg-17-945-2020},
  url = {https://bg.copernicus.org/articles/17/945/2020/},
  urldate = {2023-07-19},
  langid = {english}
}

@article{liu.etal_2019a,
  title = {Reduced Resilience as an Early Warning Signal of Forest Mortality},
  author = {Liu, Yanlan and Kumar, Mukesh and Katul, Gabriel G. and Porporato, Amilcare},
  date = {2019-11},
  journaltitle = {Nature Climate Change},
  shortjournal = {Nat. Clim. Chang.},
  volume = {9},
  number = {11},
  pages = {880--885},
  publisher = {{Nature Publishing Group}},
  issn = {1758-6798},
  doi = {10.1038/s41558-019-0583-9},
  url = {https://www.nature.com/articles/s41558-019-0583-9},
  urldate = {2023-01-10},
  issue = {11},
  langid = {english}
}

@article{lovejoy.nobre_2018,
  title = {Amazon {{Tipping Point}}},
  author = {Lovejoy, Thomas E. and Nobre, Carlos},
  date = {2018-02-21},
  journaltitle = {Science Advances},
  volume = {4},
  number = {2},
  pages = {eaat2340},
  publisher = {{American Association for the Advancement of Science}},
  doi = {10.1126/sciadv.aat2340},
  url = {https://www.science.org/doi/full/10.1126/sciadv.aat2340},
  urldate = {2023-07-18}
}

@article{lovejoy.nobre_2019,
  title = {Amazon Tipping Point: {{Last}} Chance for Action},
  shorttitle = {Amazon Tipping Point},
  author = {Lovejoy, Thomas E. and Nobre, Carlos},
  date = {2019-12-20},
  journaltitle = {Science Advances},
  volume = {5},
  number = {12},
  pages = {eaba2949},
  publisher = {{American Association for the Advancement of Science}},
  doi = {10.1126/sciadv.aba2949},
  url = {https://www.science.org/doi/full/10.1126/sciadv.aba2949},
  urldate = {2023-07-18}
}

@article{martinez-ramos.etal_1988,
  title = {Treefall {{Age Determination}} and {{Gap Dynamics}} in a {{Tropical Forest}}},
  author = {Martinez-Ramos, Miguel and Alvarez-Buylla, Elena and Sarukhan, Jose and Pinero, Daniel},
  date = {1988},
  journaltitle = {Journal of Ecology},
  volume = {76},
  number = {3},
  eprint = {2260568},
  eprinttype = {jstor},
  pages = {700--716},
  publisher = {{[Wiley, British Ecological Society]}},
  issn = {0022-0477},
  doi = {10.2307/2260568},
  url = {https://www.jstor.org/stable/2260568},
  urldate = {2023-07-19}
}

@article{mitchard_2018,
  title = {The Tropical Forest Carbon Cycle and Climate Change},
  author = {Mitchard, Edward T. A.},
  date = {2018-07},
  journaltitle = {Nature},
  volume = {559},
  number = {7715},
  pages = {527--534},
  publisher = {{Nature Publishing Group}},
  issn = {1476-4687},
  doi = {10.1038/s41586-018-0300-2},
  url = {https://www.nature.com/articles/s41586-018-0300-2},
  urldate = {2023-07-18},
  issue = {7715},
  langid = {english}
}

@online{morr.boers_2023a,
  title = {Detection of Approaching Critical Transitions in Natural Systems Driven by Red Noise},
  author = {Morr, Andreas and Boers, Niklas},
  date = {2023-10-09},
  eprint = {2310.05587},
  eprinttype = {arxiv},
  eprintclass = {physics},
  doi = {10.48550/arXiv.2310.05587},
  url = {http://arxiv.org/abs/2310.05587},
  urldate = {2023-10-13},
  pubstate = {preprint}
}

@article{negron-juarez.etal_2010,
  title = {Widespread {{Amazon}} Forest Tree Mortality from a Single Cross-Basin Squall Line Event},
  author = {Negrón-Juárez, Robinson I. and Chambers, Jeffrey Q. and Guimaraes, Giuliano and Zeng, Hongcheng and Raupp, Carlos F. M. and Marra, Daniel M. and Ribeiro, Gabriel H. P. M. and Saatchi, Sassan S. and Nelson, Bruce W. and Higuchi, Niro},
  date = {2010},
  journaltitle = {Geophysical Research Letters},
  volume = {37},
  number = {16},
  issn = {1944-8007},
  doi = {10.1029/2010GL043733},
  url = {https://onlinelibrary.wiley.com/doi/abs/10.1029/2010GL043733},
  urldate = {2022-11-30},
  langid = {english}
}

@article{negron-juarez.etal_2018,
  title = {Vulnerability of {{Amazon}} Forests to Storm-Driven Tree Mortality},
  author = {Negrón-Juárez, Robinson I. and Holm, Jennifer A. and Marra, Daniel Magnabosco and Rifai, Sami W. and Riley, William J. and Chambers, Jeffrey Q. and Koven, Charles D. and Knox, Ryan G. and McGroddy, Megan E. and Vittorio, Alan V. Di and Urquiza-Muñoz, Jose and Tello-Espinoza, Rodil and Muñoz, Waldemar Alegria and Ribeiro, Gabriel H. P. M. and Higuchi, Niro},
  date = {2018-05},
  journaltitle = {Environmental Research Letters},
  shortjournal = {Environ. Res. Lett.},
  volume = {13},
  number = {5},
  pages = {054021},
  publisher = {{IOP Publishing}},
  issn = {1748-9326},
  doi = {10.1088/1748-9326/aabe9f},
  url = {https://dx.doi.org/10.1088/1748-9326/aabe9f},
  urldate = {2022-12-02},
  langid = {english}
}

@article{newbery.etal_2011,
  title = {Growth Responses of Understorey Trees to Drought Perturbation in Tropical Rainforest in {{Borneo}}},
  author = {Newbery, D. M. and Lingenfelder, M. and Poltz, K. F. and Ong, R. C. and Ridsdale, C. E.},
  date = {2011-12-15},
  journaltitle = {Forest Ecology and Management},
  shortjournal = {Forest Ecology and Management},
  volume = {262},
  number = {12},
  pages = {2095--2107},
  issn = {0378-1127},
  doi = {10.1016/j.foreco.2011.07.030},
  url = {https://www.sciencedirect.com/science/article/pii/S0378112711004695},
  urldate = {2023-09-30}
}

@article{nicoletti.etal_2023,
  title = {The Emergence of Scale-Free Fires in {{Australia}}},
  author = {Nicoletti, Giorgio and Saravia, Leonardo and Momo, Fernando and Maritan, Amos and Suweis, Samir},
  date = {2023-03-17},
  journaltitle = {iScience},
  shortjournal = {iScience},
  volume = {26},
  number = {3},
  pages = {106181},
  issn = {2589-0042},
  doi = {10.1016/j.isci.2023.106181},
  url = {https://www.sciencedirect.com/science/article/pii/S2589004223002584},
  urldate = {2023-04-05},
  langid = {english}
}

@article{nobre.etal_2016,
  title = {Land-Use and Climate Change Risks in the {{Amazon}} and the Need of a Novel Sustainable Development Paradigm},
  author = {Nobre, Carlos A. and Sampaio, Gilvan and Borma, Laura S. and Castilla-Rubio, Juan Carlos and Silva, José S. and Cardoso, Manoel},
  date = {2016-09-27},
  journaltitle = {Proceedings of the National Academy of Sciences},
  shortjournal = {Proc. Natl. Acad. Sci. U.S.A.},
  volume = {113},
  number = {39},
  pages = {10759--10768},
  issn = {0027-8424, 1091-6490},
  doi = {10.1073/pnas.1605516113},
  url = {https://pnas.org/doi/full/10.1073/pnas.1605516113},
  urldate = {2023-07-18},
  langid = {english}
}

@article{parry.etal_2022,
  title = {Evidence of Localised {{Amazon}} Rainforest Dieback in {{CMIP6}} Models},
  author = {Parry, Isobel M. and Ritchie, Paul D. L. and Cox, Peter M.},
  date = {2022-11-24},
  journaltitle = {Earth System Dynamics},
  volume = {13},
  number = {4},
  pages = {1667--1675},
  publisher = {{Copernicus GmbH}},
  issn = {2190-4979},
  doi = {10.5194/esd-13-1667-2022},
  url = {https://esd.copernicus.org/articles/13/1667/2022/},
  urldate = {2022-12-02},
  langid = {english}
}

@article{reis.etal_2022,
  title = {Forest Disturbance and Growth Processes Are Reflected in the Geographical Distribution of Large Canopy Gaps across the {{Brazilian Amazon}}},
  author = {Reis, Cristiano Rodrigues and Jackson, Toby D. and Gorgens, Eric Bastos and Dalagnol, Ricardo and Jucker, Tommaso and Nunes, Matheus Henrique and Ometto, Jean Pierre and Aragão, Luiz E. O. C. and Rodriguez, Luiz Carlos Estraviz and Coomes, David A.},
  date = {2022},
  journaltitle = {Journal of Ecology},
  volume = {n/a},
  number = {n/a},
  issn = {1365-2745},
  doi = {10.1111/1365-2745.14003},
  url = {https://onlinelibrary.wiley.com/doi/abs/10.1111/1365-2745.14003},
  urldate = {2022-11-30},
  langid = {english}
}

@article{rietkerk.etal_2021,
  title = {Evasion of Tipping in Complex Systems through Spatial Pattern Formation},
  author = {Rietkerk, Max and Bastiaansen, Robbin and Banerjee, Swarnendu and family=Koppel, given=Johan, prefix=van de, useprefix=true and Baudena, Mara and Doelman, Arjen},
  date = {2021-10-08},
  journaltitle = {Science},
  volume = {374},
  number = {6564},
  pages = {eabj0359},
  publisher = {{American Association for the Advancement of Science}},
  doi = {10.1126/science.abj0359},
  url = {https://www.science.org/doi/full/10.1126/science.abj0359},
  urldate = {2023-10-12}
}

@article{saatchi.etal_2021,
  title = {Detecting Vulnerability of Humid Tropical Forests to Multiple Stressors},
  author = {Saatchi, Sassan and Longo, Marcos and Xu, Liang and Yang, Yan and Abe, Hitofumi and André, Michel and Aukema, Juliann E. and Carvalhais, Nuno and Cadillo-Quiroz, Hinsby and Cerbu, Gillian Ann and Chernela, Janet M. and Covey, Kristofer and Sánchez-Clavijo, Lina María and Cubillos, Isai V. and Davies, Stuart J. and De Sy, Veronique and De Vleeschouwer, Francois and Duque, Alvaro and Sybille Durieux, Alice Marie and De Avila Fernandes, Kátia and Fernandez, Luis E. and Gammino, Victoria and Garrity, Dennis P. and Gibbs, David A. and Gibbon, Lucy and Gowae, Gae Yansom and Hansen, Matthew and Lee Harris, Nancy and Healey, Sean P. and Hilton, Robert G. and Johnson, Christine May and Kankeu, Richard Sufo and Laporte-Goetz, Nadine Therese and Lee, Hyongki and Lovejoy, Thomas and Lowman, Margaret and Lumbuenamo, Raymond and Malhi, Yadvinder and Albert Martinez, Jean-Michel M. and Nobre, Carlos and Pellegrini, Adam and Radachowsky, Jeremy and Román, Francisco and Russell, Diane and Sheil, Douglas and Smith, Thomas B. and Spencer, Robert G. M. and Stolle, Fred and Tata, Hesti Lestari and Torres, Dennis del Castillo and Tshimanga, Raphael Muamba and Vargas, Rodrigo and Venter, Michelle and West, Joshua and Widayati, Atiek and Wilson, Sylvia N. and Brumby, Steven and Elmore, Aurora C.},
  date = {2021-07-23},
  journaltitle = {One Earth},
  shortjournal = {One Earth},
  volume = {4},
  number = {7},
  pages = {988--1003},
  issn = {2590-3322},
  doi = {10.1016/j.oneear.2021.06.002},
  url = {https://www.sciencedirect.com/science/article/pii/S2590332221003444},
  urldate = {2022-11-19},
  langid = {english}
}

@article{scheffer.etal_2009,
  title = {Early-Warning Signals for Critical Transitions},
  author = {Scheffer, Marten and Bascompte, Jordi and Brock, William A. and Brovkin, Victor and Carpenter, Stephen R. and Dakos, Vasilis and Held, Hermann and family=Nes, given=Egbert H., prefix=van, useprefix=true and Rietkerk, Max and Sugihara, George},
  date = {2009-09},
  journaltitle = {Nature},
  shortjournal = {Nature},
  volume = {461},
  number = {7260},
  pages = {53--59},
  issn = {0028-0836, 1476-4687},
  doi = {10.1038/nature08227},
  url = {http://www.nature.com/articles/nature08227},
  urldate = {2022-01-31},
  langid = {english}
}

@article{sierra.etal_2021,
  title = {The Fate and Transit Time of Carbon in a Tropical Forest},
  author = {Sierra, Carlos A. and Estupinan-Suarez, Lina M. and Chanca, Ingrid},
  date = {2021},
  journaltitle = {Journal of Ecology},
  volume = {109},
  number = {8},
  pages = {2845--2855},
  issn = {1365-2745},
  doi = {10.1111/1365-2745.13723},
  url = {https://onlinelibrary.wiley.com/doi/abs/10.1111/1365-2745.13723},
  urldate = {2023-07-19},
  langid = {english}
}

@article{smith.etal_2022,
  title = {Empirical Evidence for Recent Global Shifts in Vegetation Resilience},
  author = {Smith, Taylor and Traxl, Dominik and Boers, Niklas},
  date = {2022-04-28},
  journaltitle = {Nature Climate Change},
  shortjournal = {Nat. Clim. Chang.},
  pages = {1--8},
  publisher = {{Nature Publishing Group}},
  issn = {1758-6798},
  doi = {10.1038/s41558-022-01352-2},
  url = {https://www.nature.com/articles/s41558-022-01352-2},
  urldate = {2022-05-03},
  langid = {english}
}

@article{staal.etal_2015,
  title = {Synergistic Effects of Drought and Deforestation on the Resilience of the South-Eastern {{Amazon}} Rainforest},
  author = {Staal, Arie and Dekker, Stefan C. and Hirota, Marina and family=Nes, given=Egbert H., prefix=van, useprefix=true},
  date = {2015-06-01},
  journaltitle = {Ecological Complexity},
  shortjournal = {Ecological Complexity},
  volume = {22},
  pages = {65--75},
  issn = {1476-945X},
  doi = {10.1016/j.ecocom.2015.01.003},
  url = {https://www.sciencedirect.com/science/article/pii/S1476945X15000057},
  urldate = {2023-01-10},
  langid = {english}
}

@article{staal.etal_2020,
  title = {Hysteresis of Tropical Forests in the 21st Century},
  author = {Staal, Arie and Fetzer, Ingo and Wang-Erlandsson, Lan and Bosmans, Joyce H. C. and Dekker, Stefan C. and family=Nes, given=Egbert H., prefix=van, useprefix=true and Rockström, Johan and Tuinenburg, Obbe A.},
  date = {2020-10-05},
  journaltitle = {Nature Communications},
  shortjournal = {Nat Commun},
  volume = {11},
  number = {1},
  pages = {4978},
  publisher = {{Nature Publishing Group}},
  issn = {2041-1723},
  doi = {10.1038/s41467-020-18728-7},
  url = {https://www.nature.com/articles/s41467-020-18728-7},
  urldate = {2023-07-18},
  issue = {1},
  langid = {english}
}

@article{staver.etal_2011,
  title = {The {{Global Extent}} and {{Determinants}} of {{Savanna}} and {{Forest}} as {{Alternative Biome States}}},
  author = {Staver, A. Carla and Archibald, Sally and Levin, Simon A.},
  date = {2011-10-14},
  journaltitle = {Science},
  volume = {334},
  number = {6053},
  pages = {230--232},
  publisher = {{American Association for the Advancement of Science}},
  doi = {10.1126/science.1210465},
  url = {https://www.science.org/doi/full/10.1126/science.1210465},
  urldate = {2022-11-19}
}

@article{stocker.etal_2019,
  title = {Drought Impacts on Terrestrial Primary Production Underestimated by Satellite Monitoring},
  author = {Stocker, Benjamin D. and Zscheischler, Jakob and Keenan, Trevor F. and Prentice, I. Colin and Seneviratne, Sonia I. and Peñuelas, Josep},
  date = {2019-04},
  journaltitle = {Nature Geoscience},
  shortjournal = {Nat. Geosci.},
  volume = {12},
  number = {4},
  pages = {264--270},
  publisher = {{Nature Publishing Group}},
  issn = {1752-0908},
  doi = {10.1038/s41561-019-0318-6},
  url = {https://www.nature.com/articles/s41561-019-0318-6},
  urldate = {2023-05-02},
  issue = {4},
  langid = {english}
}

@article{taubert.etal_2018,
  title = {Global Patterns of Tropical Forest Fragmentation},
  author = {Taubert, Franziska and Fischer, Rico and Groeneveld, Jürgen and Lehmann, Sebastian and Müller, Michael S. and Rödig, Edna and Wiegand, Thorsten and Huth, Andreas},
  date = {2018-02},
  journaltitle = {Nature},
  volume = {554},
  number = {7693},
  pages = {519--522},
  publisher = {{Nature Publishing Group}},
  issn = {1476-4687},
  doi = {10.1038/nature25508},
  url = {https://www.nature.com/articles/nature25508},
  urldate = {2022-07-13},
  issue = {7693},
  langid = {english}
}

@article{vannes.etal_2014,
  title = {Tipping Points in Tropical Tree Cover: Linking Theory to Data},
  shorttitle = {Tipping Points in Tropical Tree Cover},
  author = {family=Nes, given=Egbert H., prefix=van, useprefix=true and Hirota, Marina and Holmgren, Milena and Scheffer, Marten},
  date = {2014},
  journaltitle = {Global Change Biology},
  volume = {20},
  number = {3},
  pages = {1016--1021},
  issn = {1365-2486},
  doi = {10.1111/gcb.12398},
  url = {https://onlinelibrary.wiley.com/doi/abs/10.1111/gcb.12398},
  urldate = {2022-06-16},
  langid = {english}
}

@article{vannes.scheffer_2007,
  title = {Slow {{Recovery}} from {{Perturbations}} as a {{Generic Indicator}} of a {{Nearby Catastrophic Shift}}.},
  author = {family=Nes, given=Egbert~H., prefix=van, useprefix=true and Scheffer, Marten},
  date = {2007-06},
  journaltitle = {The American Naturalist},
  volume = {169},
  number = {6},
  pages = {738--747},
  publisher = {{The University of Chicago Press}},
  issn = {0003-0147},
  doi = {10.1086/516845},
  url = {https://www.journals.uchicago.edu/doi/full/10.1086/516845},
  urldate = {2023-05-08}
}

@article{vasseur.yodzis_2004,
  title = {The {{Color}} of {{Environmental Noise}}},
  author = {Vasseur, David A. and Yodzis, Peter},
  date = {2004},
  journaltitle = {Ecology},
  volume = {85},
  number = {4},
  pages = {1146--1152},
  issn = {1939-9170},
  doi = {10.1890/02-3122},
  url = {https://onlinelibrary.wiley.com/doi/abs/10.1890/02-3122},
  urldate = {2023-09-30},
  langid = {english}
}

@article{vicente-serrano.etal_2013,
  title = {Response of Vegetation to Drought Time-Scales across Global Land Biomes},
  author = {Vicente-Serrano, Sergio M. and Gouveia, Célia and Camarero, Jesús Julio and Beguería, Santiago and Trigo, Ricardo and López-Moreno, Juan I. and Azorín-Molina, César and Pasho, Edmond and Lorenzo-Lacruz, Jorge and Revuelto, Jesús and Morán-Tejeda, Enrique and Sanchez-Lorenzo, Arturo},
  date = {2013-01-02},
  journaltitle = {Proceedings of the National Academy of Sciences},
  volume = {110},
  number = {1},
  pages = {52--57},
  publisher = {{Proceedings of the National Academy of Sciences}},
  doi = {10.1073/pnas.1207068110},
  url = {https://www.pnas.org/doi/abs/10.1073/pnas.1207068110},
  urldate = {2023-07-19}
}

@article{wissel_1984,
  title = {A Universal Law of the Characteristic Return Time near Thresholds},
  author = {Wissel, C.},
  date = {1984-12},
  journaltitle = {Oecologia},
  shortjournal = {Oecologia},
  volume = {65},
  number = {1},
  pages = {101--107},
  issn = {0029-8549, 1432-1939},
  doi = {10.1007/BF00384470},
  url = {http://link.springer.com/10.1007/BF00384470},
  urldate = {2022-01-31},
  langid = {english}
}

@article{wunderling.etal_2021,
  title = {Interacting Tipping Elements Increase Risk of Climate Domino Effects under Global Warming},
  author = {Wunderling, Nico and Donges, Jonathan F. and Kurths, Jürgen and Winkelmann, Ricarda},
  date = {2021-06-03},
  journaltitle = {Earth System Dynamics},
  shortjournal = {Earth Syst. Dynam.},
  volume = {12},
  number = {2},
  pages = {601--619},
  issn = {2190-4987},
  doi = {10.5194/esd-12-601-2021},
  url = {https://esd.copernicus.org/articles/12/601/2021/},
  urldate = {2022-01-31},
  langid = {english}
}

@article{wunderling.etal_2022c,
  title = {Recurrent Droughts Increase Risk of Cascading Tipping Events by Outpacing Adaptive Capacities in the {{Amazon}} Rainforest},
  author = {Wunderling, Nico and Staal, Arie and Sakschewski, Boris and Hirota, Marina and Tuinenburg, Obbe A. and Donges, Jonathan F. and Barbosa, Henrique M. J. and Winkelmann, Ricarda},
  date = {2022-08-09},
  journaltitle = {Proceedings of the National Academy of Sciences},
  volume = {119},
  number = {32},
  pages = {e2120777119},
  publisher = {{Proceedings of the National Academy of Sciences}},
  doi = {10.1073/pnas.2120777119},
  url = {https://www.pnas.org/doi/full/10.1073/pnas.2120777119},
  urldate = {2022-11-19}
}

@article{wunderling.etal_2023a,
  title = {Global Warming Overshoots Increase Risks of Climate Tipping Cascades in a Network Model},
  author = {Wunderling, Nico and Winkelmann, Ricarda and Rockström, Johan and Loriani, Sina and Armstrong McKay, David I. and Ritchie, Paul D. L. and Sakschewski, Boris and Donges, Jonathan F.},
  date = {2023-01},
  journaltitle = {Nature Climate Change},
  shortjournal = {Nat. Clim. Chang.},
  volume = {13},
  number = {1},
  pages = {75--82},
  publisher = {{Nature Publishing Group}},
  issn = {1758-6798},
  doi = {10.1038/s41558-022-01545-9},
  url = {https://www.nature.com/articles/s41558-022-01545-9},
  urldate = {2023-09-30},
  issue = {1},
  langid = {english}
}

@article{yang.etal_2019b,
  title = {The Predictability of Ecological Stability in a Noisy World},
  author = {Yang, Qiang and Fowler, Mike S. and Jackson, Andrew L. and Donohue, Ian},
  date = {2019-02},
  journaltitle = {Nature Ecology \& Evolution},
  shortjournal = {Nat Ecol Evol},
  volume = {3},
  number = {2},
  pages = {251--259},
  publisher = {{Nature Publishing Group}},
  issn = {2397-334X},
  doi = {10.1038/s41559-018-0794-x},
  url = {https://www.nature.com/articles/s41559-018-0794-x},
  urldate = {2023-09-30},
  issue = {2},
  langid = {english}
}

\clearpage
\appendix

\setcounter{section}{0}

\section{Double-well potential}\label{sec:dw}

We repeat our $\alpha$-stable Lévy noise experiments with a generic model based on a double-well potential. More specifically, we construct a double-well potential with similar time scales and value range as the Amazon model with small time scales introduced in the main body of this work. Let $x \in [0, 100]$ be an abstraction of the rainforest state and $c \in \mathbb{R}$ be an external forcing (think \emph{climate change}). Our model becomes the first-order ODE:
\begin{equation}\label{eq:dw}
    \frac{dx}{dt} = \frac{\tau}{a} ((\frac{x}{a} + b)^3 + \frac{x}{a} + b - \frac{4\sqrt{4}}{27}c) + dN_t
\end{equation}
With a time scale $\tau = 1000$, a value range $a = 20$ and a value shift $b = -3$ as fixed parameters. For $c < -1$, Eq.~\ref{eq:dw} has one stable fixed point (the rainforest state, $x \approx 80$), for $-1 \leq c \leq 1$, there are two stable fixed points exist and for $c > 1$ there is again only one stable fixed point exists (the savannah state $x \approx 40$). As in the main text, we choose $\alpha$-stable Lévy noise for the noise term $dN_t$.

\begin{figure}
    \centering
    \includegraphics[width=\linewidth]{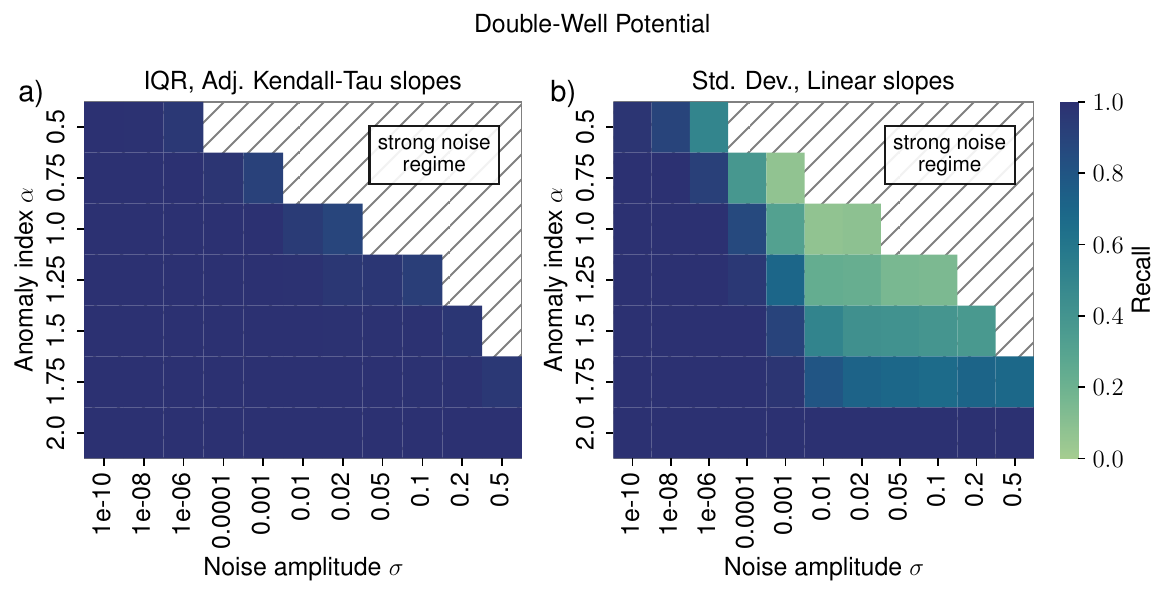} 
   \caption{Same as fig.~\ref{fig:fig3}, but using a double-well potential as the model. Recall across various noise amplitudes and anomaly indices ($\alpha = 2$ is Gaussian noise). Shown are only those noise configurations that can be considered as weak noise determined by the observed first passage time. Panel \textbf{a)} shows a processing chain that leads to high recall, i.e. the power-law disturbances are properly dealt with. Recall is close to $1.0$ for all noise configurations, with a slight decrease close to the strong noise regime. Panel \textbf{b)} in contrast shows no proper treatment of power-law disturbances, which leads to reduced recall closer to the strong noise regime.}
    \label{fig:fig3dw}
\end{figure}

These parameter settings allow us to perform simulations in the same set-up as described in Fig.~\ref{fig:fig1}, with the minor difference that we simulate climate change by linearly increasing the forcing from $c = -1.5$ to $c = 1.5$. The results are consistent with those in the main body. Fig.~\ref{fig:fig3dw} shows the recall across a variety of noise settings, for two cases: 1. adjusted Kendall-Tau slopes on the interquartile range, where recall is always high, and 2. linear slopes on the standard deviation, where some deterioration in recall can be observed towards the strong noise regime. The adjustment for a low false positive rate is less effective in the case of the double-well potential: without it, Kendall-Tau slopes on IQR get $9.3\%$ false positives, while with the adjustment, the FPR drops to $8.6\%$. Likely this is due to the threshold of $AC(1) = 0.5$ introduced in the main body, which seems valid for the Amazon rainforest model, but not for the double-well potential.

\clearpage

\section{Pink Noise}\label{sec:pink}

We generate pink noise by first generating Gaussian noise, taking its Fourier transform,  dividing each amplitude by the square root of its frequency and then transforming inversely. We then perform simulations for a range of noise amplitudes using the same Amazon rainforest models described in section~\ref{sec:methods} (both, the \citet{vannes.etal_2014} original, and our variation including small time scales). We enforce negative disturbances by mirroring the pink noise at the origin, but we keep a small additional Gaussian noise term to represent climatic variations.

The results are presented in Table~\ref{tab:tab1pink} and are consistent with the $\alpha$-stable Lévy noise results. A high recall and a low false positive rate can be achieved by using adjusted Kendall-tau slopes on the interquartile range in our model (which includes small time scales).

\begin{table}
    \centering
    \begin{tabularx}{0.75\textwidth}{Xllc@{\hspace{3em}}cc}
    \toprule
    Model & Slope & Indicator & & Recall & FPR \\
    \midrule
    van Nes & Linear & Std. Dev. & & 0.16 & 0.09 \\
    van Nes & Kendall-Tau & IQR & & 0.12 & 0.08 \\
    \midrule
    Ours & Linear & IQR & & 0.99 & 0.12 \\
     Ours & Theil-Sen & IQR & & 0.36 & 0.09 \\
     Ours & Kendall-Tau & IQR & & \textbf{1.0} & 0.11 \\
     \midrule
     Ours & Kendall-Tau & Std. Dev. & & \textbf{1.0} & 0.11 \\
     Ours & Kendall-Tau & Lag-1 AC & & \textbf{1.0} & 0.08 \\
      \midrule
      Ours & Adj. Kendall-Tau & IQR & & \textbf{1.0} & \textbf{0.0}\\
      \bottomrule
    \end{tabularx}
    \caption{Same as Table~\ref{tab:tab1}, but for pink noise. Recall and false positive rate (FPR) for various processing chains. Averages are computed across a wide choice of noise amplitudes, as long as the resulting noise can be considered weak (as measured by the first passage time). The first two rows show results for the original \citet{vannes.etal_2014} model, for which critical slowing down does not work due to a lack of representation of small time scales. The second block of rows presents different choices for computing the slope of the early warning indicator when considering our model, which includes small time scales: the non-linear Kendall-Tau slope is robust and thus preferrential. The robustness becomes clear in the third block, there is little variation in recall when comparing other early warning indicators than the interquartile-range (IQR). The final row shows that adjusting the early warning based on the absolute temporal autocorrelation reduces the FPR by five times, while keeping a high recall.}
    \label{tab:tab1pink}
\end{table}

\clearpage

\section{Linear Langevin dynamics under $\alpha$-stable noise forcing}\label{app:alphaEWS}
We have relayed the motivation behind employing variance and AC(1) as indicators for changes in system stability in the main text. This theoretical treatment regarded a linearised model under white noise forcing (see Eq. \ref{eq:langevinB}). The analytical quantities of variance and AC(1) in Eqs. \ref{eq:VarB} and \ref{eq:AC1B} directly arise from this model and approximate the expected quantities in applications to real bifurcation dynamics. The restoring rate $\lambda$ is a metric for system stability and vanishes when crossing a co-dimension 1 bifurcation. Insofar as a natural system can be viewed as being in equilibrium and experiencing small Gaussian white noise disturbances, the derived expressions dictate an increase in variance and AC(1) as the system approaches a critical transition. In the present work, we propose that Gaussian white noise forcing is an inadequate modelling choice for e.g. forest dynamics, as extreme disturbances are regularly observed. The question of whether the equilibrium dynamics of a linearised system under asymmetric, heavy-tailed noise forcing allows for a similar technique to assess CSD points to the following Langevin equation.
\begin{equation}\label{eq:langevinL}
\mathrm{d} x_t=\lambda (x_t-x^*)\mathrm{d} t+\sigma\mathrm{d} L^\alpha_t,
\end{equation}
where we assume here w.l.o.g. that $x^*=0$.
As in the main text, the noise term $\sigma\mathrm{d} L^\alpha_t$ is assumed to be $\alpha$-stable Levy-noise with $\alpha<1$, skewness parameter $\beta=-1$, and scaling parameter $c=\sigma\mathrm{d} t^{1/\alpha}$. The increments $\sigma\mathrm{d} L^\alpha_t$ are independent of each other. Their characteristic function is given by
\begin{eqnarray*}
\phi_{\sigma\mathrm{d} L^\alpha}(s)&=\exp\left(is\mu-|cs|^\alpha(1-i\beta\mathrm{sgn}(s)\tan(\pi\alpha/2))\right)\\&=\exp\left(-|\sigma\mathrm{d} t^{1/\alpha}s|^\alpha(1+i\mathrm{sgn}(s)\tan(\pi\alpha/2))\right)
\end{eqnarray*}

We will first show that there exists a stationary solution to Eq.\ref{eq:langevinL}, which is itself $\alpha$-stable.
We can write the Euler-Mayurama discretisation corresponding to Eq.\ref{eq:langevinL} as 
\begin{eqnarray}
x^{(\Delta t)}_{t+\Delta t}-x^{(\Delta t)}_t&=\lambda x^{(\Delta t)}_t\Delta t+\sigma\Delta L^\alpha_t\nonumber\\\Leftrightarrow x^{(\Delta t)}_{t+\Delta t}&=(1+\lambda\Delta t) x^{(\Delta t)}_t+\sigma\Delta L^\alpha_t\label{disc}
\end{eqnarray}
In the limit of $\Delta t\rightarrow0$, this autoregressive process converges to a solution of the continuous-time Langevin equation \ref{eq:langevinL}.
Ansatz: Assume that $x^{(\Delta t)}$ is itself $\alpha_{x^{(\Delta t)}}$-stable with parameters $\beta_{x^{(\Delta t)}}$ and $c_{x^{(\Delta t)}}$. For this to be a stationary distribution, one can see via induction on the discrete-time equation that first two the parameters must be the same as those of the noise, i.e. $\alpha_{x^{(\Delta t)}}=\alpha$ and $\beta_{x^{(\Delta t)}}=\beta=-1$. The correct choice of $c_{x^{(\Delta t)}}$ remains to be calculated. For this, notice that the two random variables on the right-hand side of \ref{disc} are independent, so the characteristic function of the total must be the product of the two individual ones:
\begin{eqnarray*}
\phi_{x^{(\Delta t)}}(s)&=\phi_{(1+\lambda\Delta t)x^{(\Delta t)}}(s)\phi_{\sigma\Delta L^\alpha}(s)\\
\Leftrightarrow \exp((-|c_{x^{(\Delta t)}}s|^\alpha&(1+i\mathrm{sgn}(s)\tan(\pi\alpha/2))))\nonumber\\
=\exp\Big(-(|(1+\lambda\Delta t)&c_{x^{(\Delta t)}}s|^\alpha+|\sigma\Delta t^{1/\alpha}s|^\alpha)(1+i\mathrm{sgn}(s)\tan(\pi\alpha/2))\Big)\\
\Leftrightarrow|c_{x^{(\Delta t)}}s|^\alpha&=|(1+\lambda\Delta t)c_{x^{(\Delta t)}}s|^\alpha+|\sigma\Delta t^{1/\alpha}s|^\alpha\\
\Leftrightarrow c_{x^{(\Delta t)}}^\alpha&=(1+\lambda\Delta t)^\alpha c_{x^{(\Delta t)}}^\alpha+\sigma^\alpha\Delta t\\
\Leftrightarrow c_{x^{(\Delta t)}}^\alpha&=\frac{\sigma^\alpha\Delta t}{1-(1+\lambda\Delta t)^\alpha}
\end{eqnarray*}
In the limit of $\Delta t\rightarrow0$, this implies that $$c_x^\alpha=\frac{-\sigma^\alpha}{\alpha\lambda}.$$
This result is consistent with the known variance $c_x^2=\frac{-\sigma^2}{2\lambda}$ for the Gaussian white noise case of $\alpha=2$.

Since the scaling parameter $c_x$ dictates the distribution width of the observed process, we may posit a direct influence of the linear restoring rate $\lambda$ on the observed variance and the IQR, analogous to the case of Gaussian white noise. We have to bear in mind however, that in this setting of an unbounded observable, the variance of the $\alpha$-stable distribution is ill-defined and will numerically diverge to infinity when applying law of large numbers estimators. In the application of a bounded forest state variable, we expect the finite variance to be an increasing function of $\lambda$, which can thus function as a CSD indicator. Further analysis is needed to theoretically motivate the use of AC(1) in a similar fashion. However, as has been laid out during the analysis within the main text, its use is warranted on a numerical basis.

\end{document}